\documentclass[twocolumn]{aastex631}
\usepackage{enumitem}

\begin{document}

\newcommand{\Fe}{\ion{Fe}{1}~6302~\AA}
\newcommand{\NaD}{\ion{Na}{1}~D1~5896~\AA}
\newcommand{\ca}{\ion{Ca}{2}~8542~\AA}

\newcommand{\Blos}{\ensuremath{B_\mathrm{LOS}}}





\title{Multiline Spectropolarimetric Evidence for Photospheric Reconnection in Ellerman Bombs from DKIST}


\author{Rahul Yadav}
\affiliation{Laboratory for Atmospheric and Space Physics, University of Colorado, Boulder, CO 80303, USA; \textnormal{rahul.yadav@lasp.colorado.edu}}

\author{Maria D. Kazachenko}
\affiliation{Laboratory for Atmospheric and Space Physics, University of Colorado, Boulder, CO 80303, USA; \textnormal{maria.kazachenko@lasp.colorado.edu}}
\affiliation{ National Solar Observatory, 3665 Discovery Drive, 80303, Boulder, CO, USA}
\affiliation{Dept. of Astrophysical and Planetary Sciences, University of Colorado, Boulder, 2000 Colorado Ave, 80305, Boulder, CO, USA}

\author{Viggo Hansteen}
\affiliation{Rosseland Centre for Solar Physics, University of Oslo, PO Box 1029, Blindern 0315, Oslo, Norway}
\affiliation{Institute of Theoretical Astrophysics, University of Oslo, P.O. Box 1029 Blindern, N-0315 Oslo, Norway}
\affiliation{SETI Institute, 339 Bernardo Ave, Suite 200, Mountain View, CA 94043, USA}
         




\begin{abstract}
Ellerman bombs (EBs) are compact, transient brightenings observed in the wings of Balmer lines and are believed to result from magnetic reconnection in the lower solar atmosphere. However, their signatures have rarely been observed simultaneously in multiple spectral lines sensitive to the lower atmosphere.
We present novel high-resolution multiline spectropolarimetric observations of EBs obtained with the Visible Spectro-Polarimeter (ViSP) on the Daniel K. Inouye Solar Telescope (DKIST). Simultaneous observations reveal remarkable H$\beta$ wing enhancements extending up to $\sim$4~\AA\ from the line core and intensity enhancements in the \ion{Na}{1}~D1 and \ion{Fe}{1}~6302 lines.
Using multiline non-LTE inversion of the \ion{Na}{1} and \ion{Fe}{1} lines, we infer localized photospheric heating of up to $\Delta T\sim$800~K, accompanied by weak upflows and downflows at EB locations. 
Comparing with synthetic observables from Bifrost radiative-MHD simulations, we find that synthetic spectra reproduce the observed H$\beta$ and \ion{Na}{1}~D wing enhancements and yield atmospheric stratification broadly consistent with those inferred from the non-LTE inversion. 
This study provides strong evidence that EBs are driven by photospheric magnetic reconnection below the temperature-minimum region, and places new constraints on the realistic EB models and small-scale energy release in the lower solar atmosphere. 
 \end{abstract}

\section{Introduction}
Magnetic reconnection is a fundamental process in astrophysical plasma that releases stored magnetic energy into other forms, such as heating and plasma flows. On the Sun, reconnection manifests itself across a wide range of spatial and temporal scales, from the largest flares to the smallest transient brightenings \citep{2022LRSP...19....1P}.

One of the manifestations of magnetic reconnection is small-scale, intense, short-lived brightenings observed in the wings of Balmer and other chromospheric lines \citep{2011ApJ...736...71W,2013ApJ...779..125N, 2015ApJ...812...11V, 2023A&A...673A..11R,2025ApJ...985..157D, 2025A&A...700A.214L,2026ApJ..1006...51C}. \cite{1917ApJ....46..298E} first described these compact brightenings as ``Hydrogen bombs'' in the wings of Balmer lines, now commonly known as Ellerman Bombs (EBs). These events are characterized by strong wing brightenings that remain invisible in the line cores, suggesting that they form in the upper photosphere or lower chromosphere.


Observations show that EBs are compact ($<$2\arcsec) and short-lived, typically lasting only a few minutes to less than an hour. They are commonly observed in emerging-flux, active, and also quiet-Sun (QS) regions \citep{2016A&A...592A.100R, 2020A&A...641L...5J}, usually near mixed-polarity magnetic fields (for reviews, see \citealt{2013JPhCS.440a2007R,2018SSRv..214..120Y}). EBs are also occasionally observed co-spatially with ultraviolet bursts (or ``IRIS bombs''), small-scale heating events in the upper chromosphere and transition region \citep{Peter2014Sci}. 

Several mechanisms have been proposed to explain the formation of EBs, including magnetic reconnection between emerging and preexisting flux \citep{2008ApJ...684..736W, 2019A&A...632A.112Y, 2023ApJ...958...54Y}, reconnection between shearing unipolar fields \citep{2002ApJ...575..506G}, and the resistive emergence of undulatory flux tubes \citep{2004ApJ...614.1099P} in the lower solar atmosphere. 

Recent 3D radiative-MHD simulations have successfully modeled EB formation from the upper convection zone to the lower solar atmosphere, including realistic radiative transfer effects. These studies show that EB-like brightenings are produced by magnetic reconnection between opposite-polarity magnetic fields in emerging-flux regions in the lower solar atmosphere \citep{2017A&A...601A.122D,2017ApJ...839...22H,2019A&A...626A..33H}.
 
Although EBs have been studied for more than a century, their thermal and magnetic structures remain poorly constrained. Previous observations were limited by narrow-band imaging, restricted spectral coverage, low cadence, or insufficient spatial resolution to resolve the small-scale magnetic structures associated with EBs \citep{1997A&A...323..599D,2013ApJ...779..125N,2015ApJ...808..133R,2023A&A...673A..11R}. Moreover, simultaneous diagnostics of the photosphere and lower chromosphere, together with direct measurements of the magnetic field at the reconnection sites, remain a challenge. The high spatial resolution, high spectral resolution, and multiline spectropolarimetric capabilities of the Daniel K. Inouye Solar Telescope (DKIST; \citealt{2020SoPh..295..172R}) now allow detailed investigations of the lower solar atmosphere. 

In this study, we present high-resolution multiline spectropolarimetric observations of EBs obtained with the Visible Spectro-Polarimeter (ViSP; \citealt{2022SoPh..297...22D}) at the DKIST. Simultaneous observations of H$\beta$, \ion{Na}{1}~D1, and \ion{Fe}{1} provide new constraints on the thermal and magnetic structure of EBs. By combining non-local thermodynamic equilibrium (non-LTE) inversions with synthetic observables derived from 3D radiative-MHD Bifrost simulations, we investigate the signatures of magnetic reconnection in the lower solar atmosphere. The paper is organized as follows. Section~\ref{Sec:observations} describes observations and data reduction. The methodology and results are presented in Sections~\ref{sec:method_and_data} and \ref{sec:results}, respectively, followed by the discussion and conclusions in Sections~\ref{sec:Discussion} and \ref{sec:conclusion}.



\section{Data Acquisition, Reduction and Overview of Observations}
\label{Sec:observations}
\begin{figure*}[!ht]
    \centering
        \includegraphics[clip,trim=0.35cm 0.25cm 0.2cm 0.25cm,width=1\linewidth]{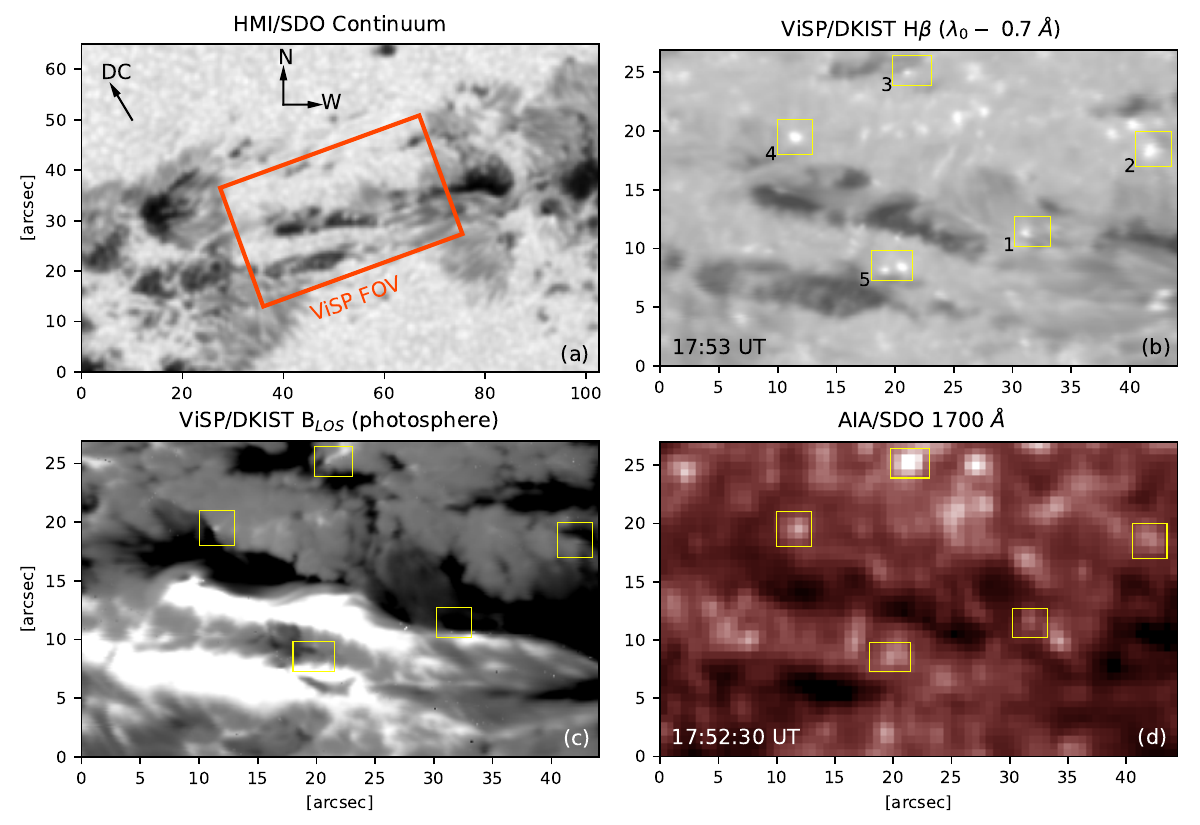}
    \caption{Overview of the field of view (FOV) observed on 18 June 2024. \textbf{(a)}: HMI/SDO continuum image providing the observational context, where the red box outlines the ViSP FOV. \textbf{(b)} ViSP H$\beta$ intensity image in the blue wing. \textbf{(c)} Line-of-sight  magnetic field inferred from the ViSP \ion{Fe}{1}~6302 line in the photosphere, where white and black represent positive and negative polarities, respectively; the magnetic field is saturated at $\pm$800~G. \textbf{(d)} AIA 1700~\AA\ image co-aligned with the ViSP FOV. The yellow boxes indicate the locations of a few example Ellerman bombs identified in the H$\beta$ wing and AIA 1700~\AA\ images, together with their associated magnetic field configurations.}
    \label{fig:overview}
\end{figure*}
\subsection{ViSP/DKIST Data Acquisition and Reduction}
In this study, we used data obtained with the ViSP instrument at the 4-m aperture DKIST, currently the largest solar telescope in the world. Under very good-seeing conditions, on 18 June 2024, from 17:53 to 18:18 UT, the ViSP instrument recorded an emerging-flux region identified in the NOAA AR 13712. The center of the ViSP field of view (FOV) was positioned at the helioprojective coordinates X = $294$\arcsec\ and Y = $-416$\arcsec\ (N13E37). The DKIST proposal ID of this dataset is pid\_2\_12.

The ViSP simultaneously captured data from three distinct spectral regions using separate detectors: the H$\beta$ 4861~\AA\ line with arm 1, the \ion{Fe}{1} 6302~\AA\ line pair with arm 2, and the \ion{Na}{1} D1 5896~\AA\ line  with arm 3. Spectral pixel sampling for ViSP arms 1 ( H$\beta$), 2 (\ion{Fe}{1}) and 3 (\ion{Na}{1}) are 8.6~m\AA, 11~m\AA, and 12~m\AA, respectively. For our observations, the width of the slit was set at $\sim$0.12\arcsec. Since the pixel scale along the slit was finer than the scan step size for all arms, we rebinned the pixels along the slit to improve the signal-to-noise ratio and to obtain square pixels of 0.12\arcsec. Spatial scanning proceeded perpendicular to the slit direction in 224 discrete steps.

The co-aligned FOV of all three lines corresponds to 26\arcsec$\times$45\arcsec.
At each scan step, full-Stokes polarization data were collected using four modulation cycles, each comprising ten modulation states. This setup led to an integration time of 1.5 seconds per slit position, which resulted in a raster cadence of $\sim$6.3 minutes. 

The ViSP calibration pipelines\footnote{https://docs.dkist.nso.edu/projects/visp/en/v5.4.1/index.html} convert Level-0 data from the telescope into Level-1 science data, which include initial wavelength calibration, corrections for instrumental
effects, polarimetric calibration, and normalization of the spectral profiles to the average disk-center quiet-Sun spectrum. For our scientific analysis, we further refined the wavelength calibration and corrected for residual polarization cross-talk following the procedure described by \citet{2024ApJ...973L..10Y}.


\subsection{Overview of Observations}
Figure~\ref{fig:overview} presents an overview of the ViSP/DKIST observations together with the corresponding contextual data from the Helioseismic and Magnetic Imager (HMI; \citealt{2012SoPh..275..207S}) and the Atmospheric Imaging Assembly (AIA; \citealt{2012SoPh..275...17L}) onboard the Solar Dynamics Observatory (SDO; \citealt{2012SoPh..275....3P}). 
Bright compact features are clearly visible in the blue wing of H$\beta$ (e.g., see Figure~\ref{fig:overview}b) indicating the presence of EBs. For these features, ViSP H$\beta$ spectra also exhibit wing enhancements (`moustache'-like shape), confirming that they are EBs. The yellow boxes in Figure~\ref{fig:overview} mark some examples of EB locations together with their associated magnetic field configurations and their counterpart in the AIA 1700~\AA\ channel.

Using the ViSP/DKIST, the line-of-sight (LOS) magnetic field inferred from the \ion{Fe}{1}~6302 line (see Section~\ref{sec:method_and_data}) reveals that these brightenings are associated with small-scale magnetic structures. Although most EBs are located at sites of opposite magnetic polarity, some events are also detected in apparently unipolar regions (e.g., box 1 in Figure~\ref{fig:overview}). 
EBs are also visible in the AIA 1600~\AA\ and 1700~\AA\ channels. However, because the 1600~\AA\ channel contains a larger contribution from transition-region emission and may include signatures of UV bursts or microflares \citep{2019ApJ...870..114S,2019A&A...626A...4V}, we primarily use the AIA 1700~\AA\ channel for context (Figure~\ref{fig:overview}d).
The co-aligned AIA 1700~\AA\ image shows enhanced emission at locations corresponding to the H$\beta$ brightenings. This enhancement is consistent with heating in the upper photosphere/lower solar atmosphere, where the AIA 1700~\AA\ passband is primarily sensitive, although chromospheric spectral lines may also contribute during transient heating events \citep{2019ApJ...870..114S}.

Due to the relatively low spatial resolution of AIA/SDO ($\sim$1.5\arcsec), most of the small-scale EBs observed in the 1700~\AA\ channel appear as diffuse brightenings, whereas they are compact, bright and well resolved in the ViSP observations (e.g., see box 5 in Figure~\ref{fig:overview}). Similarly, the high-resolution ViSP magnetograms reveal clear mixed-polarity magnetic fields at several EB locations that are not resolved in the HMI magnetograms.

\section{Methods}\label{sec:method_and_data}
\subsection{Bifrost Simulation and Spectral Synthesis}
We use the Bifrost radiative MHD code to model an emerging-flux region where multiple reconnection events occur throughout the 3D domain. 
The code solves the MHD equations for a plasma with ionization state in LTE. The computational domain extends from 2.5~Mm below the photosphere to 14.5~Mm above in the vertical direction, and spans 24~Mm in both horizontal directions. The simulation domain is resolved by $768 \times 768 \times 768$ grid points, corresponding to a horizontal grid spacing of 31.25~km. In the vertical direction, the grid spacing is nonuniform, with a resolution of approximately 12~km between 1.0~Mm below the photosphere and 4.5~Mm into the chromosphere. Additional details on the simulation setup are provided in \cite{2019A&A...626A..33H}.

To compare the simulations with the observations, we synthesized the ViSP spectral lines (\ion{Fe}{1}, \ion{Na}{1}~D1, and H$\beta$) from the Bifrost model atmosphere using the synthesis module of STiC, which is based on the RH radiative transfer code \citep{2001ApJ...557..389U}. The \ion{Fe}{1} line pair  was synthesized under the assumption of LTE, while the \ion{Na}{1}~D1 and H$\beta$ lines were treated with non-LTE.
The accurate synthesis of the hydrogen line ideally requires a full 3D treatment of radiative transfer, as 3D effects significantly influence the formation of the line core \citep{2012ApJ...749..136L}. However, since Ellerman bombs primarily form in the upper photosphere and lower chromosphere, the non-LTE synthesis under the 1.5D approximation is generally sufficient to reproduce the H$\beta$ wings reasonably well, as demonstrated in \cite{2019A&A...626A..33H}. Therefore, we adopted the simplified non-LTE 1.5D approach for the synthesis of the H$\beta$ line as the full 3D radiative transfer calculations for H$\beta$ remain computationally challenging. 

To facilitate a direct comparison with the ViSP observations, we degraded the synthetic spectra by convolving them with a Gaussian kernel of varying full width at half maximum (FWHM). By comparing the power spectra of the degraded Bifrost maps with those of the ViSP observations, we found that a Gaussian kernel with an FWHM of 4 pixels ($\sim$0.17\arcsec) provides the closest match. To account for instrumental stray light, we further convolved the spectra with a Lorentzian kernel of FWHM 10 pixels ($\sim$0.45\arcsec). Unlike a Gaussian profile, the Lorentzian profile has extended wings that better reproduce the effects of stray light.

We also compared the intensity contrasts of the degraded synthetic maps with those observed by ViSP. After applying the combined Gaussian and Lorentzian degradations, the synthetic maps show reasonable agreement with the observations in both spatial power distribution and intensity contrast. The Bifrost model atmosphere and resulting intensity maps are further discussed in Section~\ref{sec:bifrost_comparison}.

\subsection{Inversion of the ViSP Spectra}
\label{subsec:inversion}
To infer the photospheric magnetic field vector, we applied the SPIN Milne–Eddington inversion code \citep{2017SoPh..292..105Y} to the ViSP \ion{Fe}{1}~6302.5~\AA\ spectral line. We derived the stratification of atmospheric parameters within the emerging-flux region, including temperature, magnetic field strength, LOS velocity, and microturbulent velocity using the multi-line non-LTE inversion code STiC \citep{2019A&A...623A..74D}. The STiC is based on a modified implementation of the RH radiative transfer code \citep{Uitenbroek2001}, which determines atomic level populations under the assumptions of statistical equilibrium and 1.5D plane-parallel geometry. The equation of state follows the implementation used in the Spectroscopy Made Easy (SME) package \citep{2017A&A...597A..16P}, while the radiative transfer equation is solved with cubic B\'ezier formal solvers \citep{2013ApJ...764...33D}.

We used the full Stokes profiles of the \ion{Na}{1} and \ion{Fe}{1} line pairs for the STiC inversion. 
As shown in Appendix~\ref{sec:responsefunction}, the \ion{Fe}{1} lines are primarily sensitive to the photosphere, whereas the \ion{Na}{1}~D1 line probes both the photosphere and the upper photosphere/lower chromosphere.
Although ViSP observations of H$\beta$ are available, we excluded this line from the final inversions because our initial attempts to simultaneously invert H$\beta$, Na~I, and Fe~I using STiC did not satisfactorily reproduce the observed profiles. The reasons for this discrepancy remain unclear but may be related to the complex formation of H$\beta$ or limitations of the adopted 1.5D radiative transfer treatment. Although 3D effects are important for H$\alpha$ formation \citep{2012ApJ...749..136L}, their significance for H$\beta$, particularly during EB events, remains uncertain. Therefore, we restricted our final inversions to the \ion{Na}{1} and \ion{Fe}{1} lines. For the inversion, we synthesized the \ion{Fe}{1} lines in LTE under the assumption of complete frequency redistribution, whereas the \ion{Na}{1}~D1 line was synthesized under non-LTE conditions with partial redistribution
effects of scattered photons.
To accelerate inversions, we used a machine-learning-trained model atmosphere as an initial guess, improving the inversion speed by approximately 30--50\% compared to conventional random initialization \citep{2025ApJ...989..183Y}.

\section{Results}
\begin{figure*}[!ht]
    \centering
    \includegraphics[clip,trim=0.45cm 0.1cm 0.6cm 0.2cm,width=1\linewidth]{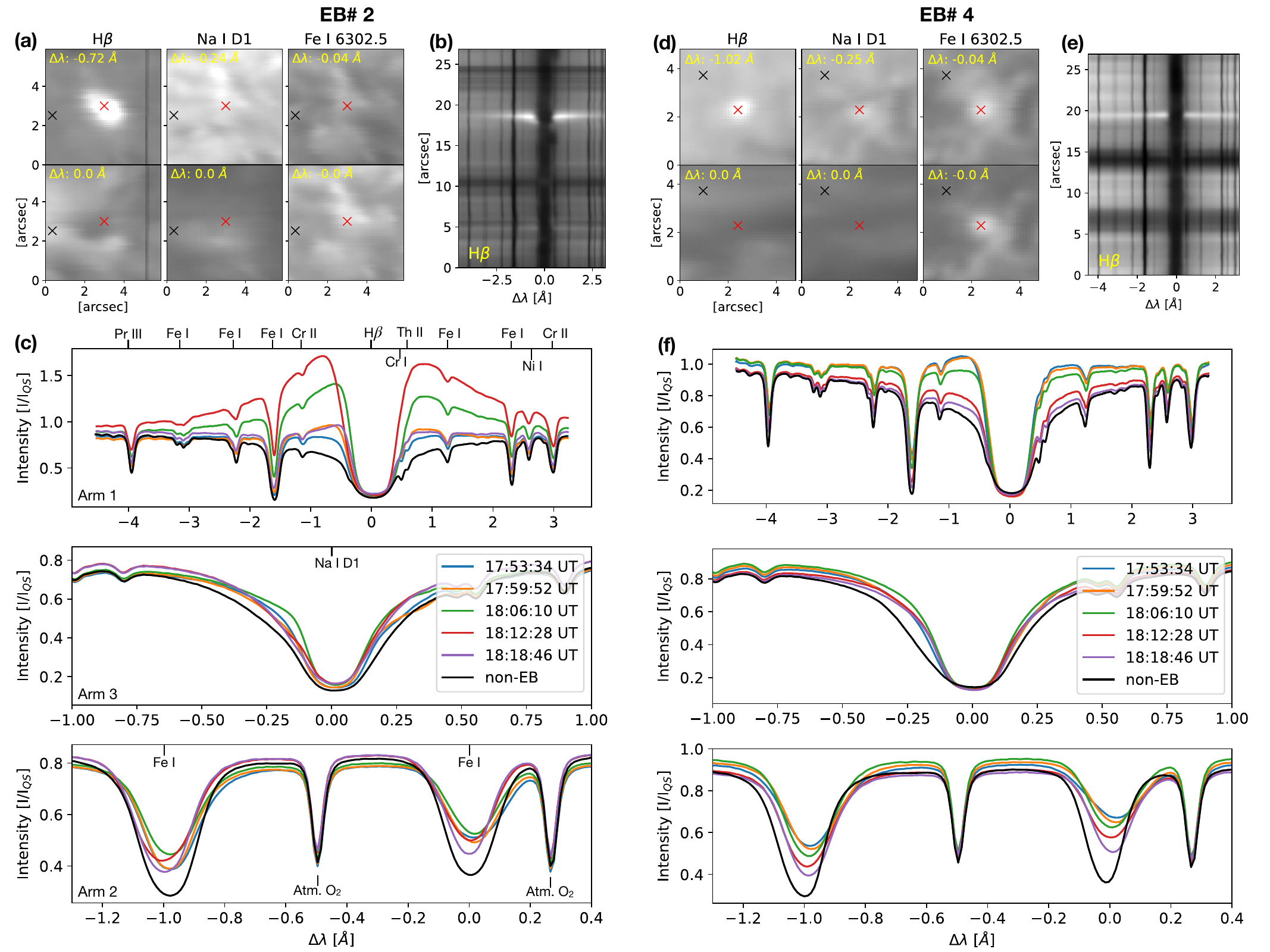}
    \caption{Multiline appearance of EBs in the DKIST/ViSP observations (H$\beta$, \ion{Na}{1}, and \ion{Fe}{1} lines). Left column (a, b and c): \textbf{(a)} images of the EB\# 2, corresponding to box 2 in Figure~\ref{fig:overview}, shown in the blue wings and line cores of different spectral lines. Red and blue crosses mark EB and non-EB locations. \textbf{(b)} Vertical spectral cut across the EB in H$\beta$, where the y-axis represents the spatial position along the slit and the x-axis denotes wavelength, illustrating `moustaches' signature of the EB. \textbf{(c)} Temporal evolution of the EB spectra in three ViSP arms at the location marked by the red cross in the top panels. The black profiles represent non-EB spectra at a pixel away from the EB site, marked as the black cross symbol in the top panels. Right panels (d, e and f): same as left panels, but for EB\# 4, outlined with box 4 in Figure~\ref{fig:overview}.}
    \label{fig:EB1}
\end{figure*}
\label{sec:results}
\subsection{Spectral Signatures of two Ellerman Bombs in H$\beta$, \ion{Na}{1}, and \ion{Fe}{1} lines}
\label{sec:spectral_signature}
Our simultaneous multiline (multi-height) observations from ViSP, combined with its high spatial and spectral resolution, enable us to investigate EB signatures in the \ion{Na}{1}, \ion{Fe}{1}, and H$\beta$ lines, which are sensitive to the photosphere and lower chromosphere where EBs are believed to form.


As an example, Figure~\ref{fig:EB1} shows the spectral appearance and temporal evolution of the intensity profiles of the EBs observed in the H$\beta$, \ion{Na}{1}, and \ion{Fe}{1} lines. EBs are remarkably visible as compact brightenings in the wings of H$\beta$, while no enhancement is present in the line core. This behavior is further illustrated in the wavelength–spatial cut across the EB, where the wing emission is strongly enhanced whereas the line core remains unaffected. We also note that the wing enhancement in H$\beta$ notably extends up to $\sim$4~\AA\ from the line core. 

For the EB$\#$2 shown in the left panels of Figure~\ref{fig:EB1}, the maximum contrast ratio at a given wavelength position, defined as 
$I_{\mathrm{ctr}} =(I_{\mathrm{EB},\lambda} -I_{\mathrm{QS},\lambda})/I_{\mathrm{QS},\lambda}$, reaches 185\% in the wings of H$\beta$, which is similar to previous findings (see the models presented in Figure~4 of \citealt{2014A&A...567A.110B} and \citealt{2019A&A...626A...4V}).
The EB also exhibits enhanced wing emission in the \ion{Na}{1} line, with contrast values of approximately 40\%, although the enhancement is weaker than in H$\beta$. In contrast to the strong wing enhancement, the \ion{Na}{1} line core shows only small contrast variations ($I_{\mathrm{ctr}} < 10\%$). Consistent with our observations, \citet{1997A&A...323..599D} reported intensity enhancements in the inner wings of the \ion{Na}{1}~D2 line associated with EB-like features observed in narrow-band filtergrams.

For the  EB$\#$4 shown in the Figure~\ref{fig:EB1}, which is relatively weaker, the \ion{Na}{1} core remains essentially unaffected. However, clear wing enhancements are observed in both the H$\beta$ and \ion{Na}{1} lines.
In contrast, the \ion{Fe}{1} line shows a significant increase in core intensity, with $I_{\mathrm{ctr}}$ values reaching $\sim$70\%. This behavior is consistent with previous observations \citep{2013ApJ...779..125N,2023MNRAS.521.3882K}, although those studies were limited by the lower spectral resolution.




\subsection{Stratified Model Atmosphere of two Ellerman Bombs from Multiline Inversion}
\label{sec:inversion}

\begin{figure*}
    \centering
        \includegraphics[clip,trim=.1cm 0.cm 0.cm 0.8cm,width=0.48\linewidth]{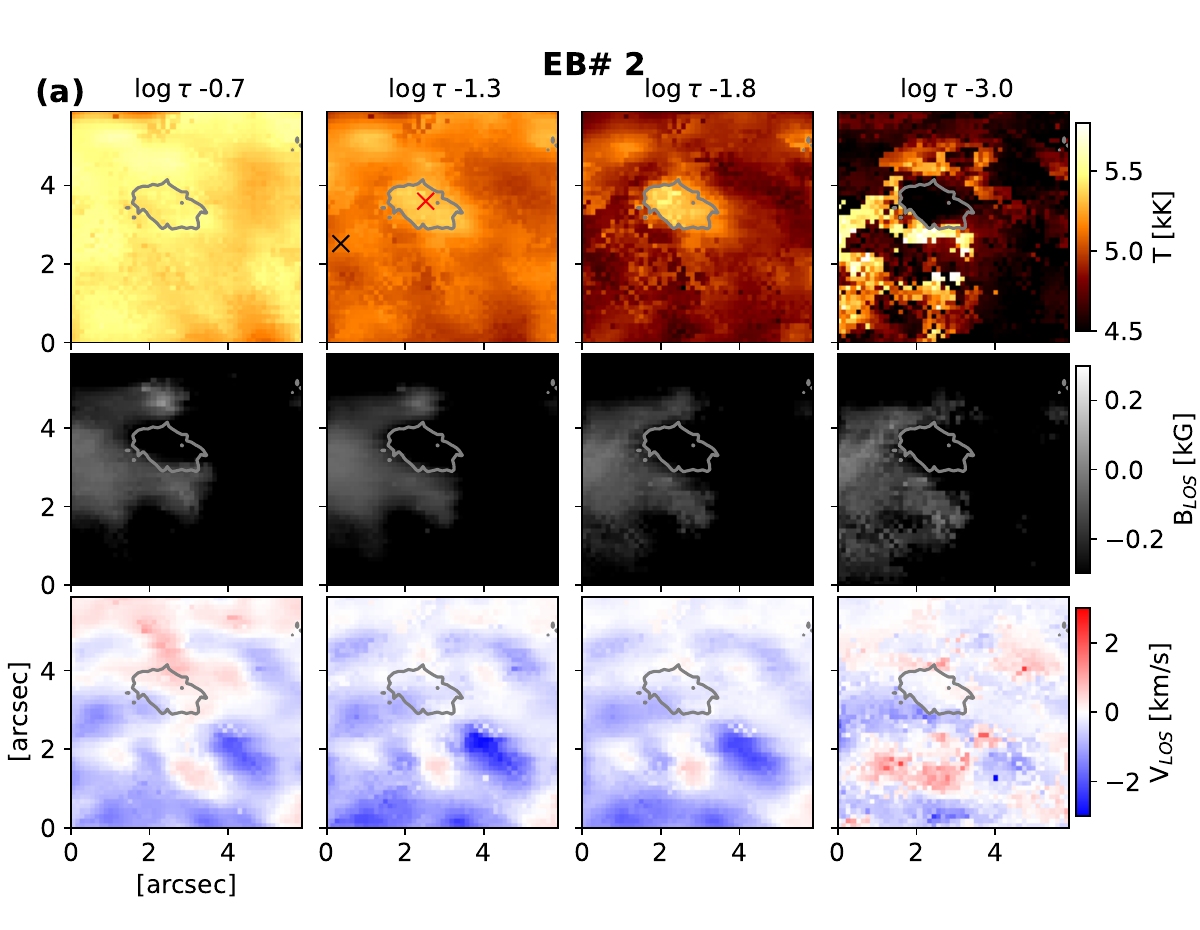}
        \includegraphics[clip,trim=0.1cm 0.cm 0.cm 0.8cm,width=0.48\linewidth]{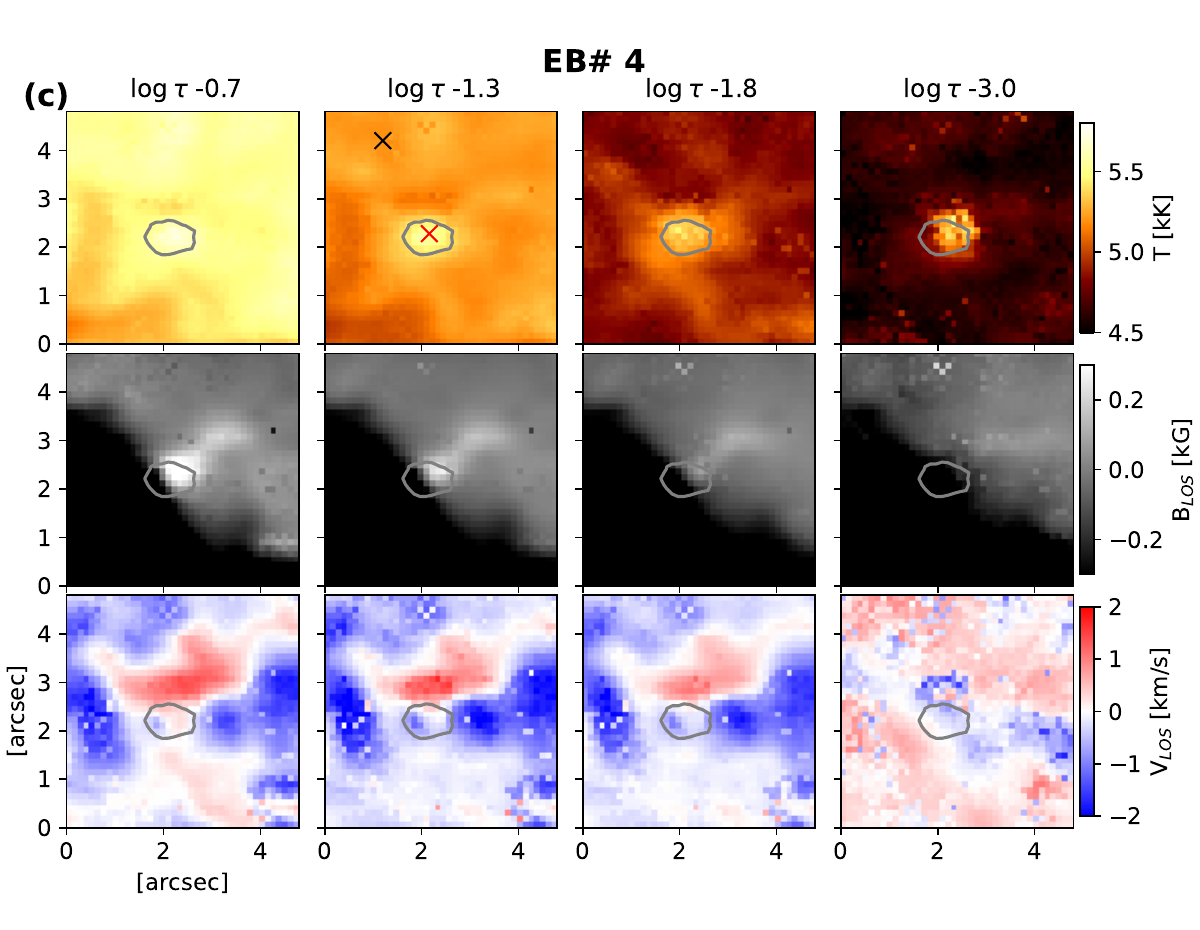}
        \includegraphics[clip,trim=.0cm 0.2cm 0.cm 0.4cm,width=0.48\linewidth]{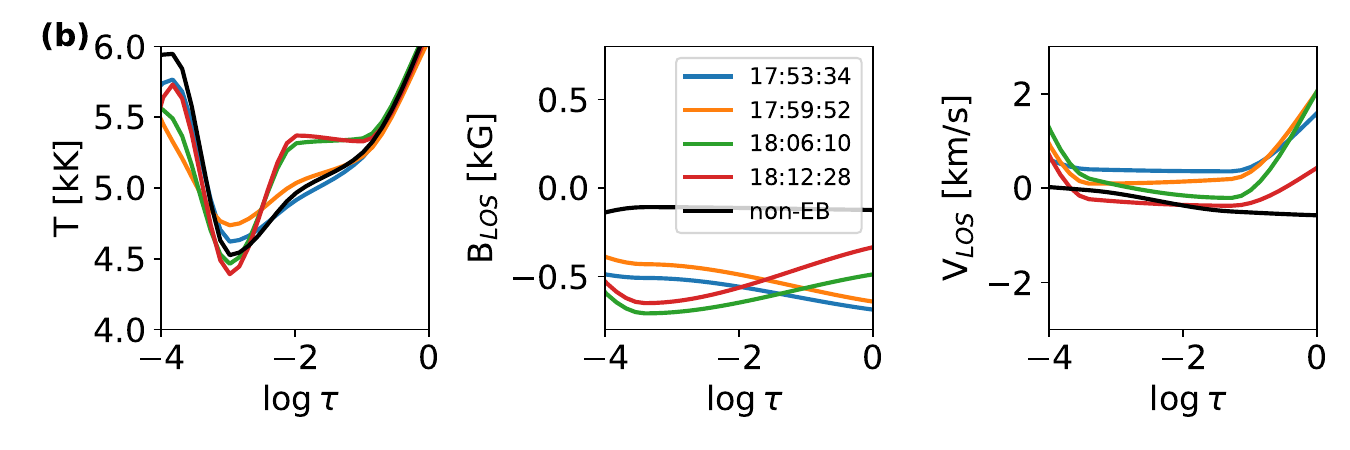}
        \includegraphics[clip,trim=.0cm 0.2cm 0.cm 0.4cm,width=0.48\linewidth]{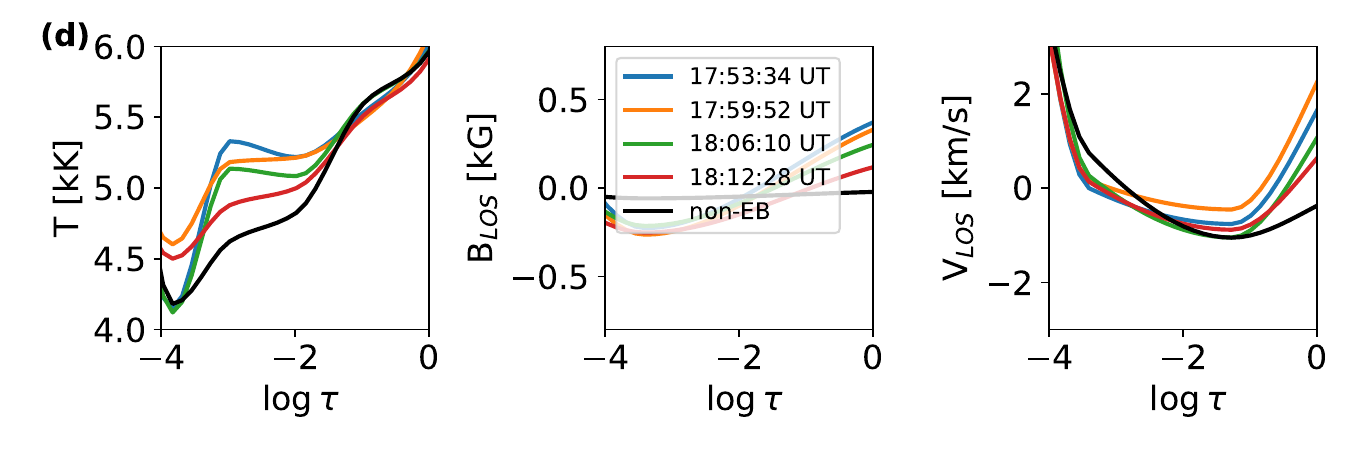}

        \caption{Stratification of physical parameters inferred from ViSP multiline inversion. Left panels: (a) shows the retrieved atmospheric parameters (temperature, LOS magnetic field, and LOS velocity) at different optical depths for EB$\#$~2 highlighted in Figure~\ref{fig:overview} and \ref{fig:EB1}. The gray contours outline the heated EB location. Panel (b) displays the temporal evolution of the inferred parameters at the EB pixel marked by the red cross, whereas the black line refers to the non-EB reference pixel marked as black cross in the panel a. Right panels (c and d): same as left, but for the EB$\#$~4 highlighted in Figure~\ref{fig:overview} and \ref{fig:EB1}.}
        \label{fig:EB1_inversion}
\end{figure*}
\begin{figure*}
    \centering
    \includegraphics[clip,trim=.3cm 0.3cm 0.3cm 0.3cm,width=1
    \linewidth]{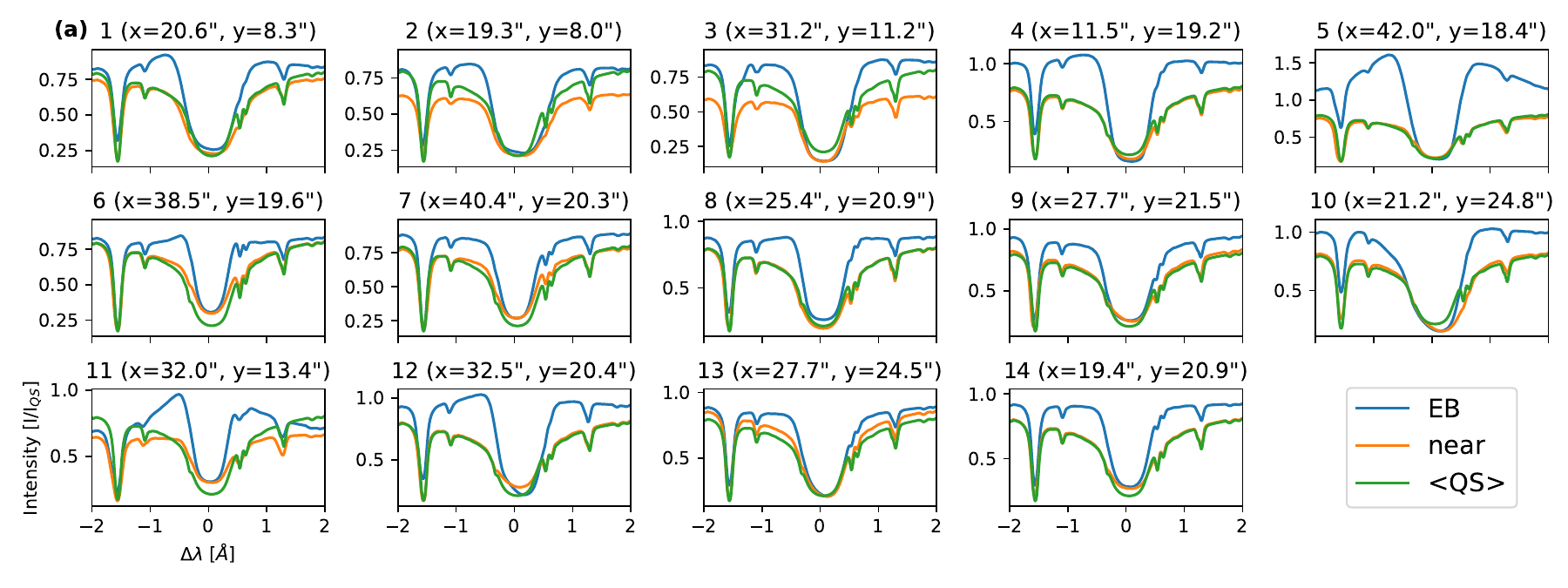}
         \includegraphics[clip,trim=.4cm .5cm 0.4cm 0.3cm,width=1\linewidth]{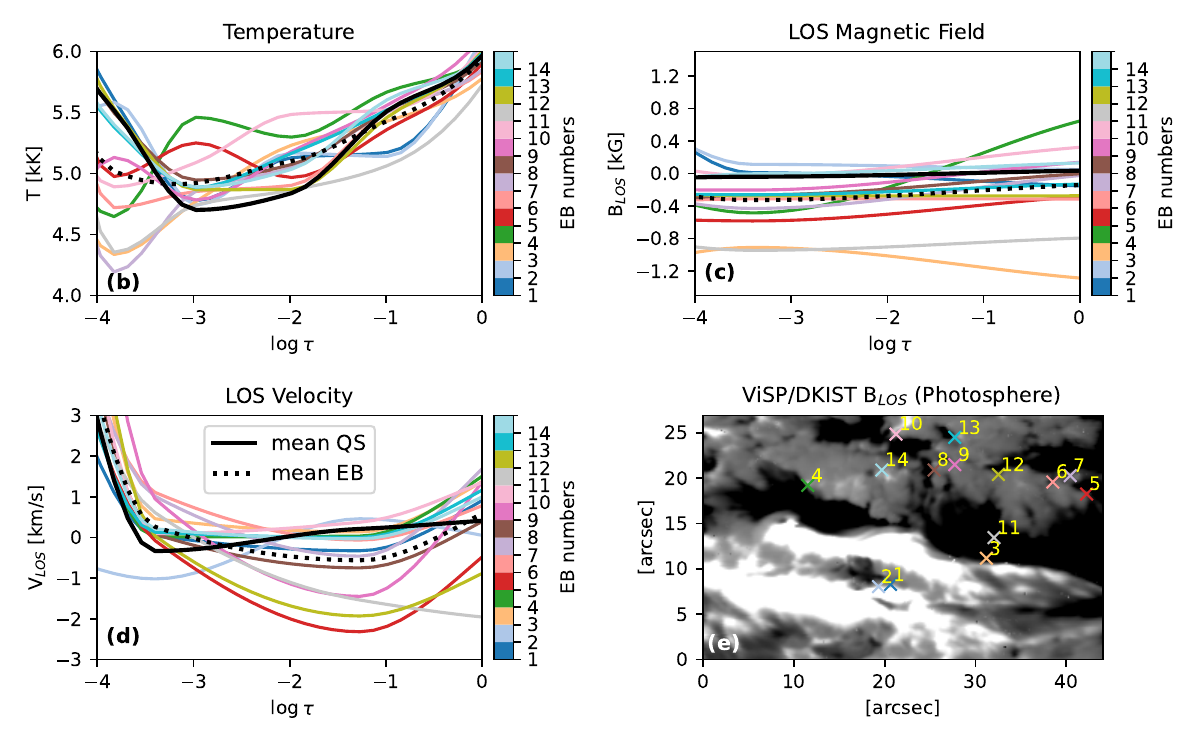}    
    \caption{Spectral profiles and inferred model atmospheres of 14 EBs. (a) H$\beta$ spectral profiles at the EB pixel (blue), a nearby non-EB pixel (orange), and the average QS profile (green). The title of each panel indicates the EB label and location. Bottom panels: inferred temperature (b), LOS magnetic field (c), and LOS velocity (d) at the EB locations. The solid and dashed black profiles represent the mean QS atmosphere and the mean atmosphere of the 14 EBs, respectively. (e) Spatial locations of all 14 unique EBs identified in the four ViSP raster maps, marked by crosses (`$\times$') on the photospheric LOS magnetic field map (saturated at $\pm$500~G), where white and black colors refer to positive and negative polarities, respectively.}
    \label{fig:EBstats}
\end{figure*}

To derive the stratified model atmosphere of the EB region, we performed simultaneous multiline inversions of the \ion{Fe}{1} and \ion{Na}{1}~D1 lines, excluding H$\beta$, using the STiC code (see Section~\ref{subsec:inversion}). The simultaneous inversion of these two spectral lines provides a more reliable reconstruction of the lower solar atmosphere (see Appendix~\ref{sec:responsefunction}).

Figure~\ref{fig:EB1_inversion} presents the resulting stratified model atmosphere for the EB region highlighted by boxes~2 and 4 in Figures~\ref{fig:overview}. The retrieved atmospheric parameters---temperature, LOS magnetic field, and LOS velocity---are displayed as functions of the logarithm of the optical depth at 500~nm (log~$\tau$). 

The temporal evolution at the EB$\#$2 exhibits a remarkably strong temperature enhancement at log~$\tau_{500}$ $\approx-2$ that reaches $\sim$800~K above that of a nearby QS reference pixel (see Figure~\ref{fig:EB1_inversion}). The maximum inferred temperature at this EB site is $\sim$5.5~kK. These temperature stratifications are in good agreement with the semi-empirical EB models reported by \cite{2014A&A...567A.110B} . 

Figure~\ref{fig:EB1_inversion}c,d present the inferred model atmosphere for EB$\#$4 highlighted by box~4 in Figure~\ref{fig:overview}. Similar to the previous example (Figure~\ref{fig:EB1_inversion}a,b), this EB also exhibits enhanced temperatures in the deeper atmosphere, although the strongest heating occurs slightly higher, around log~$\tau_{500}\approx-3$, where the temperature reaches $\sim$5.3~kK. The temperature maps reveal clear localized heating in the lower solar atmosphere situated near the PIL between two opposite-polarity magnetic concentrations. This atmospheric layer corresponds to the height range where the core of the \ion{Fe}{1} line has more sensitivity (see Appendix~\ref{sec:responsefunction}). As a consequence of heating at this height, the intensity contrast in the \ion{Fe}{1} line core increases by up to 70\% (see Figure~\ref{fig:EB1}).
As shown in Figure~\ref{fig:EB1_inversion}, this observed heating is accompanied by a temporal decrease in the LOS magnetic field strength and weak plasma flows ($\sim \pm1$~km~s$^{-1}$) in the photosphere. Together, these signatures support the interpretation that the observed EBs are driven by magnetic reconnection.



\subsection{Statistical Properties and Energetics of Ellerman Bombs}
To identify EBs in the four ViSP raster maps, we applied intensity-threshold criteria to the H$\beta$ wing maps, as described in Appendix~\ref{app:EBdetection}. Using these criteria, we identified 14 unique EBs across the four ViSP raster maps. The mean area of all identified EBs is 0.53~Mm$^2$, which corresponds to effective mean radius of 0.57\arcsec (0.41~Mm).
Figure~\ref{fig:EBstats} presents the H$\beta$ spectral profiles and inferred atmospheric properties of the 14 identified EBs. The H$\beta$ profiles show clear wing enhancements ($\pm$ 0.7~\AA) compared to both nearby non-EB pixels (located close to the EB site) and the average quiet-Sun (QS) profile, derived from QS patch (100 pixels) located within the ViSP FOV. 
Although the magnitude and shape of the enhancement vary among individual events, most EB profiles are relatively symmetric, while a few events exhibit noticeable wing asymmetries (e.g., EBs~$\#$10 and $\#$11 in Figure~\ref{fig:EBstats}a).
We also examined the intensity contrast ($I_{\rm ctr}$), as defined in Section~\ref{sec:spectral_signature}, for all 14 events. The contrast of the H$\beta$ wing ranges from 20 to 185\%, with a mean of 41\%. In the \ion{Fe}{1} line core, $I_{\rm ctr}$ ranges from 15\% to 95\%, with a mean of 51\%. In contrast, no significant intensity enhancement is observed in the core of the \ion{Na}{1} line. However, its wings show relatively weak enhancements, with $I_{\rm ctr}$ ranging from 5 to 40\%, with a mean of 11\%.

For all EBs, the inferred temperature enhancement relative to nearby non-EB pixels generally occurs between $\log\tau=-1$ and $-3$, with enhancements ranging from 150 to 800~K. The LOS velocities are relatively low ($\pm1$~km~s$^{-1}$) between $\log\tau=-1.5$ and $-3$, but show downflows and uplfows in the higher and deeper layers, respectively. This low velocity could be due to the high plasma density at the formation heights of Fe I and Na D1 lines. Events exhibiting stronger H$\beta$ wing emission generally show more pronounced temperature enhancements. The spatial distribution of the EBs further shows that most events are located in mixed-polarity magnetic regions. These results suggest that the magnitude of the heating at different heights, together with variations in the local magnetic and velocity gradients, may contribute to the diversity of the observed EB spectral signatures.

We also compared the mean atmosphere of all 14 EBs with the mean QS atmosphere (Figure~\ref{fig:EBstats}). The mean EB temperature is enhanced by $\approx300$~K at $\log\tau=-2.5$, while the mean LOS magnetic field decreases with height. However, the LOS magnetic-field stratification alone does not allow us to constrain the reconnection height, particularly given the low cadence of our observations. Around the height of the temperature enhancement ($\log\tau\approx-2.5$), the mean LOS velocity stratification shows redshifts in the higher layers and blueshifts in the deeper layers (see the black dashed profile in Figure~\ref{fig:EBstats}d). Despite the relatively small number of EBs in our sample, the co-spatial temperature enhancement near mixed polarity region and the presence of upflows and downflows in the lower atmosphere are consistent with a magnetic reconnection scenario.

We note that due to low cadence ($\approx$~6 min) each ViSP raster samples an EB at a particular stage of its evolution. Consequently, our dataset is not well suited for investigating the temporal evolution of the inferred atmospheric parameters, magnetic flux cancellation, or velocity of all individual EBs.

Our multiline inversions indicate that the typical temperature enhancement associated with the EBs ranges from approximately 150 to 800~K over optical depths of log~$\tau_{500}\approx-1$ to $-3$. Using the approach followed by  \cite{2023MNRAS.521.3882K}, we estimate the approximate thermal energy budget as: \textbf{$E_{th} = 1.5k_{B} n \Delta T \Delta V$}, where the $k_{B}$, $n$, $\Delta T$, $\Delta V$ are the Boltzmann constant, the number density, temperature enhancement, and volume of EB, respectively.
We estimated $n$, from the mean mass density ($\rho$) inferred from the STiC inversions over log~$\tau_{500}\approx-1$ to $-3$, using $n=\rho/(X m_{\rm H}+Y m_{\rm He})$, where $X=0.91$ and $Y=0.09$ represent the adopted hydrogen and helium fractions, respectively, and m$_H$ and m$_{He}$ are the masses of hydrogen and helium atoms, respectively. This yields a number density of $n\sim10^{15}$~cm$^{-3}$, which we used to estimate the EB thermal energy budget.
The EB volume was estimated assuming a characteristic size of $0.5\arcsec$, based on the ViSP observations, and a vertical extent of $\sim300$~km. Using these values, we estimate an instantaneous thermal energy of the order of $10^{24}$--$10^{25}$~erg. Due to the low cadence ($\sim6$~min) of the ViSP observations, we cannot reliably integrate the radiative losses over the lifetime of individual EBs and therefore do not estimate the total EB energy budget. A reliable estimate of the total energy release requires high-cadence observations that resolve the temporal evolution of individual EBs.
\begin{figure}
    \centering
        \includegraphics[clip,trim=.5cm 0.4cm 0.52cm 0.4cm,width=1\linewidth]{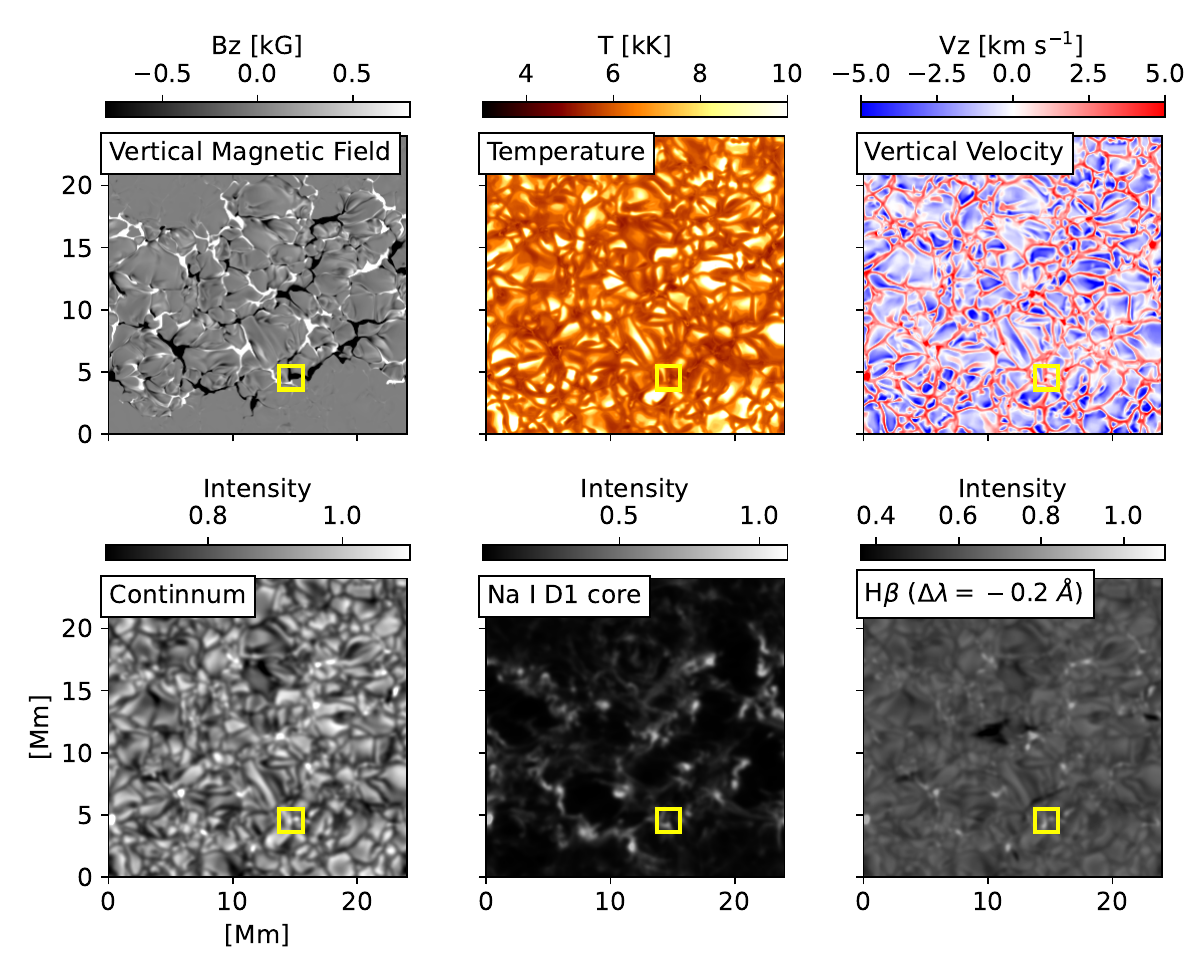}

        \includegraphics[clip,trim=.4cm 0.4cm 0.4cm 0.4cm,width=1\linewidth]{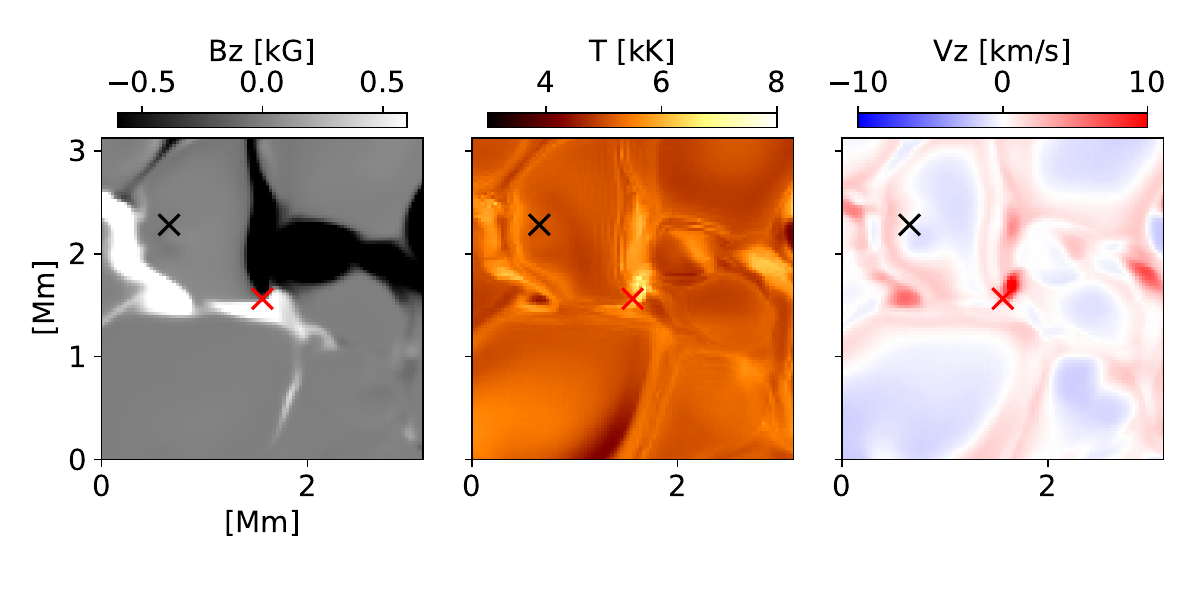}
                \includegraphics[clip,trim=.4cm 0.5cm 0.4cm 0.4cm,width=1\linewidth]{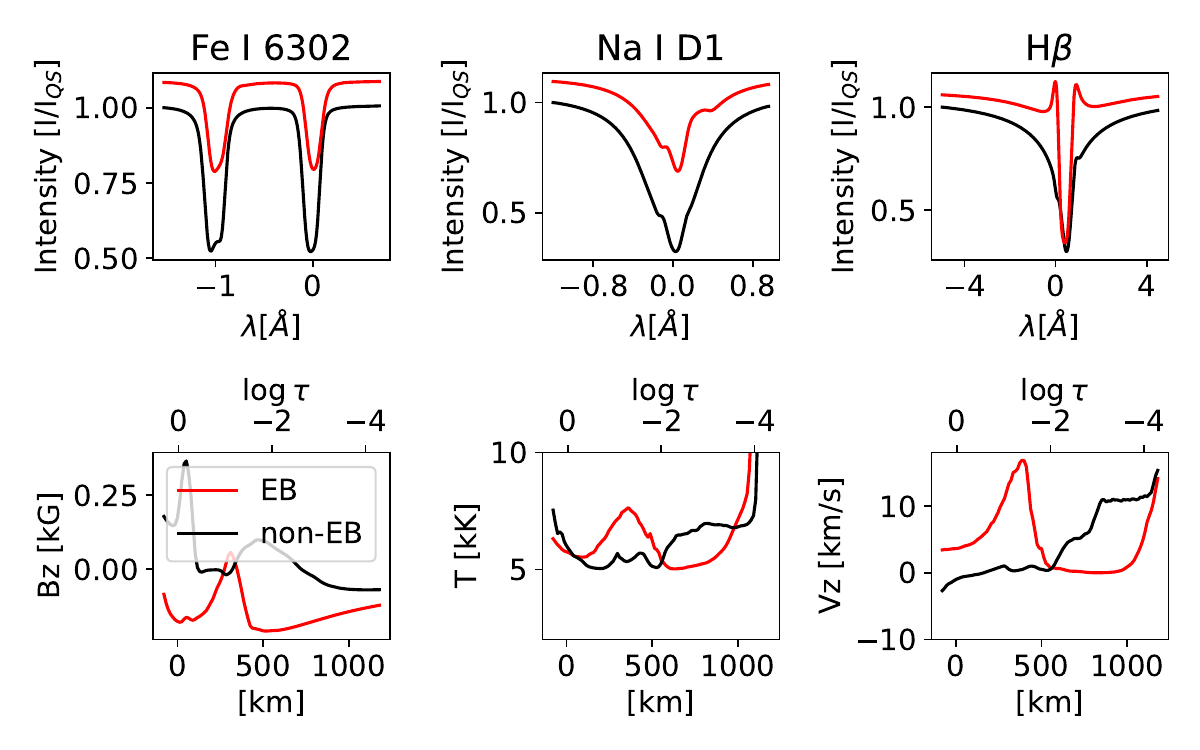}

    \caption{Bifrost model atmosphere and synthetic spectral profiles. 
    Top row: vertical component of magnetic field ($B_z$; left), where white and black refer to positive and negative polarities, respectively; temperature (\textit{T}; middle), and vertical component of velocity ($V_z$; right) at a height of 150~km above the photosphere. Second row from top: spatially-convolved synthetic images derived from the Bifrost model atmosphere. Middle row: Zoom-in for the EB region highlighted in top panels (see yellow box). Bottom two rows: Convolved synthetic spectra and stratification of magnetic field, temperature, and vertical velocity at the EB location (red cross) and a nearby reference location (black cross) marked in the middle panels.}
    \label{fig:EB1_bifrost}
\end{figure}
\subsection{Comparison of Observed and Simulated EB Spectra and Atmospheric Structures}
\label{sec:bifrost_comparison}
To facilitate a comparison with the observed spectral signatures, temperature and plasma flows derived from the ViSP observations, we selected a simulated EB whose spectral and atmospheric properties most closely resemble the observed events.
Figure~\ref{fig:EB1_bifrost} demonstrates the Bifrost model atmosphere and synthetic maps, where the selected simulated EB is shown by the yellow box.
This event is located within a compact mixed-polarity magnetic region characterized by enhanced temperatures and strong plasma flows.

The synthetic H$\beta$ profiles exhibit enhanced wing emission with the characteristic ``moustache'' structure, while the line core remains largely unaffected, reproducing the typical observational signatures of EBs. The contrast enhancement, quantified by $I_{\mathrm{ctr}}$, reaches up to $\sim$100\% in the H$\beta$ wings at $\sim$0.2~\AA\ from the line core. 

The synthetic \ion{Fe}{1}~6302 line pair and \ion{Na}{1}~D1 profiles also show significant increases in core intensity, indicating that the EB heating extends into the upper photospheric layers. The contrast values at the line cores of \ion{Fe}{1} and \ion{Na}{1}~D1 reach approximately 50\% and 90\%, respectively. The enhancement in the simulated \ion{Fe}{1} core is comparable to that observed in the ViSP data, whereas the \ion{Na}{1}~D1 core does not exhibit such strong contrast enhancements in the observations. This discrepancy may be related to the intrinsically 3D and strongly scattering nature of the \ion{Na}{1} line, effects that are not fully accounted for in our 1.5D spectral synthesis.
We also note that the continuum contrast enhancement in both the simulated and observed spectra is comparable, reaching values of approximately $\sim$10\%. 

Relative to the non-EB atmosphere, the EB atmosphere exhibits a sharp increase in temperature ($\sim$2000~K) and velocity ($\sim$15~km~s$^{-1}$), accompanied by a noticeable decrease in magnetic field strength around $\sim$350~km ($\sim$log~$\tau= -1.5$) above the photosphere (see Figure~\ref{fig:EB1_bifrost}). These signatures suggest the presence of a magnetic reconnection site and support the interpretation that EBs are driven by photospheric reconnection, leading to localized heating and dynamic plasma motions.

The observationally inferred model atmosphere also shows similar signatures of temperature enhancement, velocity perturbations, and magnetic field variations, although with smaller amplitudes (see Section~\ref{sec:inversion}). While the selected simulated EB reproduces several key observational characteristics, the simulated atmosphere is systematically hotter than the observed one. This discrepancy is likely related to the stronger magnetic fields ($>600$~G) present in the simulation domain, whereas the observed EBs often occur in regions where one of the magnetic polarities is significantly weaker, with field strengths of $\sim200$~G or less (e.g., EB$\#$4 in Figure~\ref{fig:EB1_inversion}).

Note that while we compare our observations with an existing Bifrost simulation of an Ellerman bomb (e.g., \citealt{2019A&A...626A..33H}), this simulation was not specifically designed to reproduce the conditions of our observations. Therefore it should be regarded as a qualitative comparison rather than a direct one-to-one match. A dedicated simulation with different simulation parameters may provide a better match, but such modeling is computationally expensive and beyond the scope of the present study.



\section{Discussion}
\label{sec:Discussion}
We analyzed novel high spatial, spectral and multiline spectropolarimetric observations (\ion{Fe}{1}, \ion{Na}{1}~D1, and H$\beta$) of EBs in a flux emerging region obtained with ViSP/DKIST. Although EBs are believed to form in the lower atmosphere via magnetic reconnection, the lack of simultaneous multiline observations has previously limited our ability to clearly trace their signatures in multiple lines sensitive to the lower solar atmosphere. The present observations revealed the first clear EB signatures in the H$\beta$, \ion{Na}{1}, and \ion{Fe}{1} lines, which are sensitive to the photosphere and lower chromosphere. Furthermore, the multiline inversions provided important constraints on the thermal, magnetic, and dynamical properties of EBs in the lower atmosphere. 

We analyzed 14 unique EBs identified in four ViSP raster maps. All events exhibit the characteristic enhancement in the H$\beta$ wings without a significant change in the line core, although the magnitude and shape of the wing enhancement vary considerably among individual events. The intensity contrasts further demonstrate that the EBs produce distinct signatures in spectral lines sampling different atmospheric layers. The strongest enhancements are observed in the H$\beta$ wings, with $I_{\rm ctr}$ ranging from 20\% to 185\% (mean of 41\%). Significant brightenings are also present in the \ion{Fe}{1} line core, where $I_{\rm ctr}$ ranges from 15\% to 95\% (mean of 51\%). In contrast, the \ion{Na}{1} line core shows no significant enhancement, while its wings exhibit only weak brightenings, with a mean contrast of 11\%. 
\\
Using multiline imaging spectroscopy in H$\alpha$, \ion{Na}{1}~D1, and \ion{Mg}{1}~b2, \citet{2015ApJ...808..133R} reported diffuse brightenings in these lines that appeared to be associated with preceding EBs. In comparison, the high spatial and spectral resolution of ViSP allows us to resolve clear and simultaneous EB signatures in multiple spectral lines sensitive to different heights in the lower solar atmosphere. These distinct spectral responses indicate that EB-induced heating is strongly height dependent and likely occurs in the lower solar atmosphere, below the temperature-minimum region.

The inferred temperature enhancement at the EB locations relative to the QS atmosphere generally occurs between $\log\tau=-1$ and $-3$ and ranges from approximately 150 to 800~K. Previous studies, using 1D semi-empirical modeling or non-LTE inversions, have reported EB temperature enhancements ranging from a few hundred to a few thousand kelvin, primarily between $\log\tau=-3$ and $-5$ \citep{1983SoPh...87..135K,Socas2006,2014A&A...567A.110B,2015A&A...582A.104R,2017A&A...598A..33L,2019A&A...627A.101V}. The inferred temperature enhancement at the EB locations in our study is relatively modest ($<1000$~K), which may be attributed to the relatively weak EBs in our dataset.
In addition, the LOS velocities are relatively low ($\pm1$~km~s$^{-1}$) between $\log\tau=-1$ and $-3$, although several events show weak signatures of oppositely directed flows, which is consistent with recent findings of \citealt{2026ApJ..1007L..39K}.

Averaged over all 14 EBs, the temperature is enhanced by $\approx300$~K relative to the mean QS atmosphere. Moreover, the mean LOS velocity stratification shows weak redshifts in the higher layers ($\log\tau\approx-4$) and blueshifts in the deeper layers ($\log\tau\approx-1.5$).
Most EB events are located near mixed-polarity magnetic regions, where plasma heating and upflows and downflows of plasma are observed. These observed signatures are consistent with the magnetic reconnection scenario in the lower atmosphere.
However, some EBs occur in apparently unipolar regions, which may result from unresolved opposite polarities, strongly inclined magnetic fields, or reconnection within serpentine emerging-flux systems \citep{2002ApJ...575..506G, 2002SoPh..209..119B, 2004ApJ...614.1099P}. 

Due to the low cadence ($\sim$6 min) of ViSP raster maps, our dataset is not suitable for investigating the variation of the physical parameters (e.g., magnetic flux cancellation, temperature and velocity). Of the 14 EBs, only two are detected in four consecutive raster maps. These events show a strong correlation ($r=0.95$) between the increase in temperature at $\log\tau\approx-3$ and the intensity of the H$\beta$ wings, supporting a close association between the brightening of the H $\beta$ wings and the heating of the plasma. Further investigation of their temporal relationship requires high-cadence, high-spatial-resolution multiline spectropolarimetric observations, which are now possible with DKIST.

The comparison with Bifrost radiative-MHD simulations provides further insight into the physical conditions associated with EBs. The simulations reproduce several key observational signatures, including the characteristic H$\beta$ ``moustache'' profiles and localized heating near mixed-polarity magnetic regions. However, the simulated EBs exhibit stronger heating, velocity perturbations, and magnetic-field reductions than inferred from the observations. A notable difference is that the strongest enhancement of the H$\beta$ wing occurs at $\pm0.2$~\AA\ in the simulations, compared to $\pm1$~\AA\ in the observations.

These discrepancies likely reflect limitations in the current modeling and synthesis. The Bifrost simulations do not include physical processes such as ambipolar diffusion and non-equilibrium ionization, which may influence chromospheric plasma \citep{2019A&A...626A..33H}. In addition, the spectral synthesis was performed using a 1.5D non-LTE approximation, neglecting full 3D radiative transfer effects. This limitation is particularly important for H$\beta$ and \ion{Na}{1}~D1, whose formation is strongly influenced by scattering and multidimensional radiative transfer \citep{2010ApJ...709.1362L}. Future simulations incorporating more realistic lower chromospheric physics together with full 3D radiative transfer calculations (e.g., Multi3D: \citealt{2009ASPC..415...87L} or PORTA: \citealt{2013A&A...557A.143S}) will therefore be essential for achieving a more quantitative comparison between observations and models of EBs.

\section{Conclusion}
\label{sec:conclusion}
We derive physical properties in 14 Ellerman bombs (EBs) using high-resolution spectropolarimetric observations of H$\beta$, \ion{Na}{1}~D1, and \ion{Fe}{1} lines from ViSP/DKIST, multiline inversions using \ion{Fe}{1} and \ion{Na}{1}~D1 lines, and Bifrost simulations. Our main results are summarized as follows:
\begin{itemize} [leftmargin=10pt]
    \item[--] All observed EBs show the characteristic H$\beta$ wing enhancement without significant line-core brightening. The H$\beta$ wing contrast ranges from 20\% to 185\% (mean: 41\%), while the \ion{Fe}{1} line-core contrast ranges from 15\% to 90\% (mean: 51\%). The \ion{Na}{1} line shows no significant core enhancement but exhibits weak wing brightenings with a mean contrast of 11\%.

    \item[--] Most EBs occur near mixed-polarity regions, with the exception of two events. 

    \item[--] The inferred model atmosphere from multiline inversions reveals temperature enhancements of $\sim150$ -- 800~K over $\log\tau=-1$ to $-3$ (corresponding to the upper photosphere and lower chromosphere), with a mean enhancement of $\approx300$~K. The LOS velocity stratifications are generally weak but show signatures of oppositely directed flows, with redshifts in the higher layers and blueshifts in the deeper layers.
    
    \item[--]We conclude that the derived  co-spatial heating and plasma flows, together with the location of most EBs near mixed-polarity regions, are consistent with a magnetic reconnection scenario in the lower atmosphere.

    \item[--] We find that Bifrost simulations reproduce key observed EB signatures, including H$\beta$ wing enhancement and localized heating near mixed-polarity regions. However, simulations show stronger heating, velocity perturbations, and magnetic-field changes than inferred from the ViSP observations, as well as differences in the H$\beta$ wing enhancement, highlighting the need for more realistic simulations and radiative-transfer modeling for quantitative comparisons.
\end{itemize}
Overall, our results demonstrate that DKIST observations, combined with multiline inversions and radiative MHD simulations, provide powerful diagnostics of the thermal and magnetic structure of EBs. Further progress in understanding the role of EBs in chromospheric and coronal heating, as well as their relationship to UV brightenings, will require high-cadence multiline observations spanning from the photosphere to the upper atmosphere. Coordinated observations with DKIST, IRIS \citep{2021SoPh..296...84D}, and future missions such as the Chromospheric Magnetism Explorer (CMEx; \citealt{2026Ap&SS.371...47G}) will offer unprecedented opportunities to trace how the energy released by photospheric reconnection propagates through the solar atmosphere.

\begin{acknowledgements}
We thank the anonymous referees for their constructive comments and suggestions, which have significantly improved this paper. 
We thank the US government for providing the funding that made this research possible. We acknowledge support from NASA ECIP NNH18ZDA001N and NSF CAREER SPVKK1RC2MZ3. 
We thank Gianna Cauzzi, Jo\~ao Manuel da Silva Santos, and Ivan Milic for useful discussions.
VH was also supported by NASA grant 80NSSC24K0258.
The research reported herein is based in part on data collected with the Daniel K. Inouye Solar Telescope (DKIST) a facility of the U.S. National Science Foundation.  DKIST is operated by the National Solar Observatory under a cooperative agreement with the Association of Universities for Research in Astronomy, Inc. DKIST is located on land of spiritual and cultural significance to Native Hawaiian people. The use of this important site to further scientific knowledge is done with appreciation and respect. This research has made use of NASA’s Astrophysics Data System. We acknowledge the community effort devoted to the development of the following open-source packages that were used in this work: NumPy (\url{numpy.org}), matplotlib (\url{matplotlib.org}) and SunPy (\url{sunpy.org}).

\end{acknowledgements}

\bibliographystyle{aa}
\bibliography{new-ref}  

@ARTICLE{2026ApJ..1006...51C,
       author = {{Chaurasiya}, Ravi and {Bayanna}, Ankala Raja and {Joshi}, Jayant},
        title = "{Comparative Analysis of Ellerman and Quiet Sun Ellerman Bombs in the Solar Atmosphere}",
      journal = {\apj},
         year = 2026,
        month = jul,
       volume = {1006},
       number = {1},
          eid = {51},
        pages = {51},
          doi = {10.3847/1538-4357/ae7dea},
archivePrefix = {arXiv},
       eprint = {2606.15111},
 primaryClass = {astro-ph.SR},
       adsurl = {https://ui.adsabs.harvard.edu/abs/2026ApJ..1006...51C}
}

@ARTICLE{2020A&A...641L...5J,
       author = {{Joshi}, Jayant and {Rouppe van der Voort}, Luc H.~M. and {de la Cruz Rodr{\'\i}guez}, Jaime},
        title = "{Signatures of ubiquitous magnetic reconnection in the lower solar atmosphere}",
      journal = {\aap},
         year = 2020,
        month = sep,
       volume = {641},
          eid = {L5},
        pages = {L5},
          doi = {10.1051/0004-6361/202038769},
archivePrefix = {arXiv},
       eprint = {2006.14975},
 primaryClass = {astro-ph.SR},
       adsurl = {https://ui.adsabs.harvard.edu/abs/2020A&A...641L...5J}
}

@ARTICLE{2013ApJ...779..125N,
       author = {{Nelson}, C.~J. and {Shelyag}, S. and {Mathioudakis}, M. and {Doyle}, J.~G. and {Madjarska}, M.~S. and {Uitenbroek}, H. and {Erd{\'e}lyi}, R.},
        title = "{Ellerman Bombs{\textemdash}Evidence for Magnetic Reconnection in the Lower Solar Atmosphere}",
      journal = {\apj},
         year = 2013,
        month = dec,
       volume = {779},
       number = {2},
          eid = {125},
        pages = {125},
          doi = {10.1088/0004-637X/779/2/125},
archivePrefix = {arXiv},
       eprint = {1310.7756},
 primaryClass = {astro-ph.SR},
       adsurl = {https://ui.adsabs.harvard.edu/abs/2013ApJ...779..125N}
}

@ARTICLE{2025ApJ...989..183Y,
       author = {{Yadav}, Rahul and {Kazachenko}, Maria D. and {Cauzzi}, Gianna and {Tamburri}, Cole A. and {Corchado}, Marcel and {French}, Ryan},
        title = "{Multiline Spectropolarimetric Observation of Flare Ribbon Fine Structures with ViSP/DKIST}",
      journal = {\apj},
         year = 2025,
        month = aug,
       volume = {989},
       number = {2},
          eid = {183},
        pages = {183},
          doi = {10.3847/1538-4357/adf4c1},
archivePrefix = {arXiv},
       eprint = {2507.20070},
 primaryClass = {astro-ph.SR},
       adsurl = {https://ui.adsabs.harvard.edu/abs/2025ApJ...989..183Y}
}

@ARTICLE{2023A&A...673A..11R,
       author = {{Rouppe van der Voort}, Luc H.~M. and {van Noort}, Michiel and {de la Cruz Rodr{\'\i}guez}, Jaime},
        title = "{Ultra-high-resolution observations of plasmoid-mediated magnetic reconnection in the deep solar atmosphere}",
      journal = {\aap},
         year = 2023,
        month = may,
       volume = {673},
          eid = {A11},
        pages = {A11},
          doi = {10.1051/0004-6361/202345933},
archivePrefix = {arXiv},
       eprint = {2302.11496},
 primaryClass = {astro-ph.SR},
       adsurl = {https://ui.adsabs.harvard.edu/abs/2023A&A...673A..11R}
}

@ARTICLE{2014A&A...567A.110B,
       author = {{Berlicki}, A. and {Heinzel}, P.},
        title = "{Observations and NLTE modeling of Ellerman bombs}",
      journal = {\aap},
         year = 2014,
        month = jul,
       volume = {567},
          eid = {A110},
        pages = {A110},
          doi = {10.1051/0004-6361/201323244},
archivePrefix = {arXiv},
       eprint = {1406.5702},
 primaryClass = {astro-ph.SR},
       adsurl = {https://ui.adsabs.harvard.edu/abs/2014A&A...567A.110B}
}

@ARTICLE{1983SoPh...87..135K,
       author = {{Kitai}, R.},
        title = "{On the mass motions and the atmospheric states of moustaches.}",
      journal = {\solphys},
         year = 1983,
        month = aug,
       volume = {87},
       number = {1},
        pages = {135-154},
          doi = {10.1007/BF00151165},
       adsurl = {https://ui.adsabs.harvard.edu/abs/1983SoPh...87..135K}
}

@ARTICLE{2015ApJ...808..133R,
       author = {{Rutten}, R.~J. and {Rouppe van der Voort}, L.~H.~M. and {Vissers}, G.~J.~M.},
        title = "{Ellerman Bombs at High Resolution. IV. Visibility in Na I and Mg I}",
      journal = {\apj},
         year = 2015,
        month = aug,
       volume = {808},
       number = {2},
          eid = {133},
        pages = {133},
          doi = {10.1088/0004-637X/808/2/133},
archivePrefix = {arXiv},
       eprint = {1506.04426},
 primaryClass = {astro-ph.SR},
       adsurl = {https://ui.adsabs.harvard.edu/abs/2015ApJ...808..133R}
}

@ARTICLE{2008ApJ...684..736W,
       author = {{Watanabe}, H. and {Kitai}, R. and {Okamoto}, K. and {Nishida}, K. and {Kiyohara}, J. and {Ueno}, S. and {Hagino}, M. and {Ishii}, T.~T. and {Shibata}, K.},
        title = "{Spectropolarimetric Observation of an Emerging Flux Region: Triggering Mechanisms of Ellerman Bombs}",
      journal = {\apj},
         year = 2008,
        month = sep,
       volume = {684},
       number = {1},
        pages = {736-746},
          doi = {10.1086/590234},
archivePrefix = {arXiv},
       eprint = {0805.4266},
 primaryClass = {astro-ph},
       adsurl = {https://ui.adsabs.harvard.edu/abs/2008ApJ...684..736W}
}

@ARTICLE{1917ApJ....46..298E,
       author = {{Ellerman}, Ferdinand},
        title = "{Solar Hydrogen ``bombs''}",
      journal = {\apj},
         year = 1917,
        month = nov,
       volume = {46},
        pages = {298},
          doi = {10.1086/142366},
       adsurl = {https://ui.adsabs.harvard.edu/abs/1917ApJ....46..298E}
}

@ARTICLE{2022SoPh..297...22D,
       author = {{de Wijn}, A.~G. and {Casini}, R. and {Carlile}, A. and {Lecinski}, A.~R. and {Sewell}, S. and {Zmarzly}, P. and {Eigenbrot}, A.~D. and {Beck}, C. and {W{\"o}ger}, F. and {Kn{\"o}lker}, M.},
        title = "{The Visible Spectro-Polarimeter of the Daniel K. Inouye Solar Telescope}",
      journal = {\solphys},
         year = 2022,
        month = feb,
       volume = {297},
       number = {2},
          eid = {22},
        pages = {22},
          doi = {10.1007/s11207-022-01954-1},
archivePrefix = {arXiv},
       eprint = {2203.00117},
 primaryClass = {astro-ph.IM},
       adsurl = {https://ui.adsabs.harvard.edu/abs/2022SoPh..297...22D}
}

@ARTICLE{2001ApJ...557..389U,
       author = {{Uitenbroek}, H.},
        title = "{Multilevel Radiative Transfer with Partial Frequency Redistribution}",
      journal = {\apj},
         year = 2001,
        month = aug,
       volume = {557},
       number = {1},
        pages = {389-398},
          doi = {10.1086/321659},
       adsurl = {https://ui.adsabs.harvard.edu/abs/2001ApJ...557..389U}
}

@ARTICLE{2018SSRv..214..120Y,
       author = {{Young}, Peter R. and {Tian}, Hui and {Peter}, Hardi and {Rutten}, Robert J. and {Nelson}, Chris J. and {Huang}, Zhenghua and {Schmieder}, Brigitte and {Vissers}, Gregal J.~M. and {Toriumi}, Shin and {Rouppe van der Voort}, Luc H.~M. and {Madjarska}, Maria S. and {Danilovic}, Sanja and {Berlicki}, Arkadiusz and {Chitta}, L.~P. and {Cheung}, Mark C.~M. and {Madsen}, Chad and {Reardon}, Kevin P. and {Katsukawa}, Yukio and {Heinzel}, Petr},
        title = "{Solar Ultraviolet Bursts}",
      journal = {\ssr},
         year = 2018,
        month = dec,
       volume = {214},
       number = {8},
          eid = {120},
        pages = {120},
          doi = {10.1007/s11214-018-0551-0},
archivePrefix = {arXiv},
       eprint = {1805.05850},
 primaryClass = {astro-ph.SR},
       adsurl = {https://ui.adsabs.harvard.edu/abs/2018SSRv..214..120Y}
}

@ARTICLE{2012SoPh..275....3P,
       author = {{Pesnell}, W. Dean and {Thompson}, B.~J. and {Chamberlin}, P.~C.},
        title = "{The Solar Dynamics Observatory (SDO)}",
      journal = {\solphys},
         year = 2012,
        month = jan,
       volume = {275},
       number = {1-2},
        pages = {3-15},
          doi = {10.1007/s11207-011-9841-3},
       adsurl = {https://ui.adsabs.harvard.edu/abs/2012SoPh..275....3P}
}

@ARTICLE{2012SoPh..275..207S,
   author = {{Scherrer}, P.~H. and {Schou}, J. and {Bush}, R.~I. and {Kosovichev}, A.~G. and 
	{Bogart}, R.~S. and {Hoeksema}, J.~T. and {Liu}, Y. and {Duvall}, T.~L. and 
	{Zhao}, J. and {Title}, A.~M. and {Schrijver}, C.~J. and {Tarbell}, T.~D. and 
	{Tomczyk}, S.},
    title = "{The Helioseismic and Magnetic Imager (HMI) Investigation for the Solar Dynamics Observatory (SDO)}",
  journal = {\solphys},
     year = 2012,
    month = jan,
   volume = 275,
    pages = {207-227},
      doi = {10.1007/s11207-011-9834-2},
   adsurl = {https://ui.adsabs.harvard.edu/abs/2012SoPh..275..207S}
}

@ARTICLE{2012SoPh..275...17L,
   author = {{Lemen}, J.~R. and {Title}, A.~M. and {Akin}, D.~J. and {Boerner}, P.~F. and 
	{Chou}, C. and {Drake}, J.~F. and {Duncan}, D.~W. and {Edwards}, C.~G. and 
	{Friedlaender}, F.~M. and {Heyman}, G.~F. and {Hurlburt}, N.~E. and 
	{Katz}, N.~L. and {Kushner}, G.~D. and {Levay}, M. and {Lindgren}, R.~W. and 
	{Mathur}, D.~P. and {McFeaters}, E.~L. and {Mitchell}, S. and 
	{Rehse}, R.~A. and {Schrijver}, C.~J. and {Springer}, L.~A. and 
	{Stern}, R.~A. and {Tarbell}, T.~D. and {Wuelser}, J.-P. and 
	{Wolfson}, C.~J. and {Yanari}, C. and {Bookbinder}, J.~A. and 
	{Cheimets}, P.~N. and {Caldwell}, D. and {Deluca}, E.~E. and 
	{Gates}, R. and {Golub}, L. and {Park}, S. and {Podgorski}, W.~A. and 
	{Bush}, R.~I. and {Scherrer}, P.~H. and {Gummin}, M.~A. and 
	{Smith}, P. and {Auker}, G. and {Jerram}, P. and {Pool}, P. and 
	{Soufli}, R. and {Windt}, D.~L. and {Beardsley}, S. and {Clapp}, M. and 
	{Lang}, J. and {Waltham}, N.},
    title = "{The Atmospheric Imaging Assembly (AIA) on the Solar Dynamics Observatory (SDO)}",
  journal = {\solphys},
     year = 2012,
    month = jan,
   volume = 275,
    pages = {17-40},
      doi = {10.1007/s11207-011-9776-8},
   adsurl = {https://ui.adsabs.harvard.edu/abs/2012SoPh..275...17L}
}

@ARTICLE{2021SoPh..296...84D,
       author = {{De Pontieu}, Bart and {Polito}, Vanessa and {Hansteen}, Viggo and {Testa}, Paola and {Reeves}, Katharine K. and {Antolin}, Patrick and {N{\'o}brega-Siverio}, Daniel Elias and {Kowalski}, Adam F. and {Martinez-Sykora}, Juan and {Carlsson}, Mats and {McIntosh}, Scott W. and {Liu}, Wei and {Daw}, Adrian and {Kankelborg}, Charles C.},
        title = "{A New View of the Solar Interface Region from the Interface Region Imaging Spectrograph (IRIS)}",
      journal = {\solphys},
         year = 2021,
        month = may,
       volume = {296},
       number = {5},
          eid = {84},
        pages = {84},
          doi = {10.1007/s11207-021-01826-0},
archivePrefix = {arXiv},
       eprint = {2103.16109},
 primaryClass = {astro-ph.SR},
       adsurl = {https://ui.adsabs.harvard.edu/abs/2021SoPh..296...84D}
}

@ARTICLE{2025A&A...700A.214L,
       author = {{Litwicka}, M. and {Berlicki}, A. and {Schmieder}, B.},
        title = "{Statistical analysis of compact brightenings in IRIS Mg II h and k lines}",
      journal = {\aap},
         year = 2025,
        month = aug,
       volume = {700},
          eid = {A214},
        pages = {A214},
          doi = {10.1051/0004-6361/202453134},
       adsurl = {https://ui.adsabs.harvard.edu/abs/2025A&A...700A.214L}
}

@ARTICLE{2013A&A...557A.143S,
       author = {{{\v{S}}t{\v{e}}p{\'a}n}, Ji{\v{r}}{\'\i} and {Trujillo Bueno}, Javier},
        title = "{PORTA: A three-dimensional multilevel radiative transfer code for modeling the intensity and polarization of spectral lines with massively parallel computers}",
      journal = {\aap},
         year = 2013,
        month = sep,
       volume = {557},
          eid = {A143},
        pages = {A143},
          doi = {10.1051/0004-6361/201321742},
archivePrefix = {arXiv},
       eprint = {1307.4217},
 primaryClass = {astro-ph.SR},
       adsurl = {https://ui.adsabs.harvard.edu/abs/2013A&A...557A.143S}
}

@ARTICLE{2015A&A...582A.104R,
       author = {{Rezaei}, R. and {Beck}, C.},
        title = "{Multiwavelength spectropolarimetric observations of an Ellerman bomb}",
      journal = {\aap},
         year = 2015,
        month = oct,
       volume = {582},
          eid = {A104},
        pages = {A104},
          doi = {10.1051/0004-6361/201526124},
       adsurl = {https://ui.adsabs.harvard.edu/abs/2015A&A...582A.104R}
}

@INPROCEEDINGS{2009ASPC..415...87L,
       author = {{Leenaarts}, J. and {Carlsson}, M.},
        title = "{MULTI3D: A Domain-Decomposed 3D Radiative Transfer Code}",
    booktitle = {The Second Hinode Science Meeting: Beyond Discovery-Toward Understanding},
         year = 2009,
       editor = {{Lites}, B. and {Cheung}, M. and {Magara}, T. and {Mariska}, J. and {Reeves}, K.},
       series = {Astronomical Society of the Pacific Conference Series},
       volume = {415},
        month = dec,
        pages = {87},
       adsurl = {https://ui.adsabs.harvard.edu/abs/2009ASPC..415...87L}
}

@ARTICLE{2026ApJ..1007L..39K,
       author = {{Kawabata}, Yusuke and {Katsukawa}, Yukio and {Kubo}, Masahito and {Oba}, Takayoshi and {Matsumoto}, Takuma and {Ishikawa}, Ryohtaroh T. and {Naito}, Yoshihiro and {Hara}, Hirohisa and {Shimizu}, Toshifumi and {Uraguchi}, Fumihiro and {Tsuzuki}, Toshihiro and {Shinoda}, Kazuya and {Tamura}, Tomonori and {Suematsu}, Yoshinori and {Quintero Noda}, Carlos and {del Toro Iniesta}, Jose Carlos and {Orozco Su{\'a}rez}, David and {Balaguer Jimenez}, Mar{\'\i}a and {Solanki}, Sami K. and {Lagg}, Andreas and {Gandorfer}, Achim and {Bernasconi}, Pietro and {Berkefeld}, Thomas and {Feller}, Alex and {Riethm{\"u}ller}, Tino L. and {{\'A}lvarez-Herrero}, Alberto and {Narayanamurthy}, Smitha and {Grauf}, Bianca and {Carpenter}, Michael and {Bell}, Alexander and {Mart{\'\i}nez Pillet}, Valent{\'\i}n and {Bail{\'e}n}, Francisco Javier and {Blanco Rodr{\'\i}guez}, Julian and {Castellanos Dur{\'a}n}, Juan Sebasti{\'a}n and {Harnes}, Edvarda and {Hoelken}, Johannes and {Iglesias}, Francisco A. and {Siu-Tapia}, Azaymi L. and {Strecker}, Hanna and {Vukadinovi{\'c}}, Du{\v{s}}an},
        title = "{3D Magnetic Structures of Ellerman Bombs Revealed by SUNRISE III/SCIP}",
      journal = {\apjl},
         year = 2026,
        month = aug,
       volume = {1007},
       number = {2},
          eid = {L39},
        pages = {L39},
          doi = {10.3847/2041-8213/ae91e0},
archivePrefix = {arXiv},
       eprint = {2608.00619},
 primaryClass = {astro-ph.SR},
       adsurl = {https://ui.adsabs.harvard.edu/abs/2026ApJ..1007L..39K}
}

@ARTICLE{2019A&A...623A..74D,
       author = {{de la Cruz Rodr{\'\i}guez}, J. and {Leenaarts}, J. and {Danilovic}, S. and {Uitenbroek}, H.},
        title = "{STiC: A multiatom non-LTE PRD inversion code for full-Stokes solar observations}",
      journal = {\aap},
         year = 2019,
        month = mar,
       volume = {623},
          eid = {A74},
        pages = {A74},
          doi = {10.1051/0004-6361/201834464},
archivePrefix = {arXiv},
       eprint = {1810.08441},
 primaryClass = {astro-ph.SR},
       adsurl = {https://ui.adsabs.harvard.edu/abs/2019A&A...623A..74D}
}

@ARTICLE{2002ApJ...575..506G,
       author = {{Georgoulis}, Manolis K. and {Rust}, David M. and {Bernasconi}, Pietro N. and {Schmieder}, Brigitte},
        title = "{Statistics, Morphology, and Energetics of Ellerman Bombs}",
      journal = {\apj},
         year = 2002,
        month = aug,
       volume = {575},
       number = {1},
        pages = {506-528},
          doi = {10.1086/341195},
       adsurl = {https://ui.adsabs.harvard.edu/abs/2002ApJ...575..506G}
}

@ARTICLE{2019A&A...627A.101V,
       author = {{Vissers}, G.~J.~M. and {de la Cruz Rodr{\'\i}guez}, J. and {Libbrecht}, T. and {Rouppe van der Voort}, L.~H.~M. and {Scharmer}, G.~B. and {Carlsson}, M.},
        title = "{Dissecting bombs and bursts: non-LTE inversions of low-atmosphere reconnection in SST and IRIS observations}",
      journal = {\aap},
         year = 2019,
        month = jul,
       volume = {627},
          eid = {A101},
        pages = {A101},
          doi = {10.1051/0004-6361/201833560},
archivePrefix = {arXiv},
       eprint = {1905.02035},
 primaryClass = {astro-ph.SR},
       adsurl = {https://ui.adsabs.harvard.edu/abs/2019A&A...627A.101V}
}

@ARTICLE{2017ApJ...839...22H,
       author = {{Hansteen}, V.~H. and {Archontis}, V. and {Pereira}, T.~M.~D. and {Carlsson}, M. and {Rouppe van der Voort}, L. and {Leenaarts}, J.},
        title = "{Bombs and Flares at the Surface and Lower Atmosphere of the Sun}",
      journal = {\apj},
         year = 2017,
        month = apr,
       volume = {839},
       number = {1},
          eid = {22},
        pages = {22},
          doi = {10.3847/1538-4357/aa6844},
archivePrefix = {arXiv},
       eprint = {1704.02872},
 primaryClass = {astro-ph.SR},
       adsurl = {https://ui.adsabs.harvard.edu/abs/2017ApJ...839...22H}
}

@ARTICLE{2011ApJ...736...71W,
       author = {{Watanabe}, Hiroko and {Vissers}, Gregal and {Kitai}, Reizaburo and {Rouppe van der Voort}, Luc and {Rutten}, Robert J.},
        title = "{Ellerman Bombs at High Resolution. I. Morphological Evidence for Photospheric Reconnection}",
      journal = {\apj},
         year = 2011,
        month = jul,
       volume = {736},
       number = {1},
          eid = {71},
        pages = {71},
          doi = {10.1088/0004-637X/736/1/71},
archivePrefix = {arXiv},
       eprint = {1105.4008},
 primaryClass = {astro-ph.SR},
       adsurl = {https://ui.adsabs.harvard.edu/abs/2011ApJ...736...71W}
}

@ARTICLE{2017A&A...601A.122D,
       author = {{Danilovic}, S.},
        title = "{Simulating Ellerman bomb-like events}",
      journal = {\aap},
         year = 2017,
        month = may,
       volume = {601},
          eid = {A122},
        pages = {A122},
          doi = {10.1051/0004-6361/201730403},
archivePrefix = {arXiv},
       eprint = {1701.02112},
 primaryClass = {astro-ph.SR},
       adsurl = {https://ui.adsabs.harvard.edu/abs/2017A&A...601A.122D}
}

@ARTICLE{2019A&A...626A...4V,
       author = {{Vissers}, Gregal J.~M. and {Rouppe van der Voort}, Luc H.~M. and {Rutten}, Robert J.},
        title = "{Automating Ellerman bomb detection in ultraviolet continua}",
      journal = {\aap},
         year = 2019,
        month = jun,
       volume = {626},
          eid = {A4},
        pages = {A4},
          doi = {10.1051/0004-6361/201834811},
archivePrefix = {arXiv},
       eprint = {1901.07975},
 primaryClass = {astro-ph.SR},
       adsurl = {https://ui.adsabs.harvard.edu/abs/2019A&A...626A...4V}
}

@article{2017SoPh..292..105Y,
	Adsurl = {https://ui.adsabs.harvard.edu/abs/2017SoPh..292..105Y},
	Author = {{Yadav}, Rahul and {Mathew}, Shibu K. and {Tiwary}, Alok Ranjan},
	Doi = {10.1007/s11207-017-1131-2},
	Eid = {105},
	Journal = {\solphys},
	Month = {Aug},
	Number = {8},
	Pages = {105},
	Title = {{SPIN: An Inversion Code for the Photospheric Spectral Line}},
	Volume = {292},
	Year = {2017}}

@ARTICLE{2012ApJ...749..136L,
       author = {{Leenaarts}, J. and {Carlsson}, M. and {Rouppe van der Voort}, L.},
        title = "{The Formation of the H{\ensuremath{\alpha}} Line in the Solar Chromosphere}",
      journal = {\apj},
         year = 2012,
        month = apr,
       volume = {749},
       number = {2},
          eid = {136},
        pages = {136},
          doi = {10.1088/0004-637X/749/2/136},
archivePrefix = {arXiv},
       eprint = {1202.1926},
 primaryClass = {astro-ph.SR},
       adsurl = {https://ui.adsabs.harvard.edu/abs/2012ApJ...749..136L}
}

@INPROCEEDINGS{2013JPhCS.440a2007R,
       author = {{Rutten}, Robert J. and {Vissers}, Gregal J.~M. and {Rouppe van der Voort}, Luc H.~M. and {S{\"u}tterlin}, Peter and {Vitas}, Nikola},
        title = "{Ellerman bombs: fallacies, fads, usage}",
    booktitle = {Journal of Physics Conference Series},
         year = 2013,
       series = {Journal of Physics Conference Series},
       volume = {440},
        month = jun,
    publisher = {IOP},
          eid = {012007},
        pages = {012007},
          doi = {10.1088/1742-6596/440/1/012007},
archivePrefix = {arXiv},
       eprint = {1304.1364},
 primaryClass = {astro-ph.SR},
       adsurl = {https://ui.adsabs.harvard.edu/abs/2013JPhCS.440a2007R}
}

@ARTICLE{2025ApJ...985..157D,
       author = {{da Silva Santos}, J.~M. and {Dunnington}, E. and {Jarolim}, R. and {Danilovic}, S. and {Criscuoli}, S.},
        title = "{Magnetic Reconnection in a Compact Magnetic Dome: Chromospheric Emissions and High-velocity Plasma Flows}",
      journal = {\apj},
         year = 2025,
        month = jun,
       volume = {985},
       number = {2},
          eid = {157},
        pages = {157},
          doi = {10.3847/1538-4357/adcf23},
archivePrefix = {arXiv},
       eprint = {2502.04292},
 primaryClass = {astro-ph.SR},
       adsurl = {https://ui.adsabs.harvard.edu/abs/2025ApJ...985..157D}
}

@ARTICLE{1997A&A...323..599D,
       author = {{Denker}, C.},
        title = "{Two-dimensional speckle spectroscopy of the moustache phenomenon on the Sun.}",
      journal = {\aap},
         year = 1997,
        month = jul,
       volume = {323},
        pages = {599-605},
       adsurl = {https://ui.adsabs.harvard.edu/abs/1997A&A...323..599D}
}

@ARTICLE{2002SoPh..209..119B,
       author = {{Bernasconi}, P.~N. and {Rust}, D.~M. and {Georgoulis}, M.~K. and {Labonte}, B.~J.},
        title = "{Moving Dipolar Features in an Emerging Flux Region}",
      journal = {\solphys},
         year = 2002,
        month = sep,
       volume = {209},
       number = {1},
        pages = {119-139},
          doi = {10.1023/A:1020943816174},
       adsurl = {https://ui.adsabs.harvard.edu/abs/2002SoPh..209..119B}
}

@ARTICLE{2016A&A...592A.100R,
       author = {{Rouppe van der Voort}, Luc H.~M. and {Rutten}, Robert J. and {Vissers}, Gregal J.~M.},
        title = "{Reconnection brightenings in the quiet solar photosphere}",
      journal = {\aap},
         year = 2016,
        month = aug,
       volume = {592},
          eid = {A100},
        pages = {A100},
          doi = {10.1051/0004-6361/201628889},
archivePrefix = {arXiv},
       eprint = {1606.03675},
 primaryClass = {astro-ph.SR},
       adsurl = {https://ui.adsabs.harvard.edu/abs/2016A&A...592A.100R}
}

@ARTICLE{2026Ap&SS.371...47G,
       author = {{Gilbert}, Holly and {Centeno}, Rebecca and {de Wijn}, Alfred and {Bryans}, Paul},
        title = "{Enabling critical science through a new mission: the chromospheric magnetism explorer}",
      journal = {\apss},
         year = 2026,
        month = may,
       volume = {371},
       number = {5},
          eid = {47},
        pages = {47},
          doi = {10.1007/s10509-026-04574-4},
       adsurl = {https://ui.adsabs.harvard.edu/abs/2026Ap&SS.371...47G}
}

@ARTICLE{2022LRSP...19....1P,
       author = {{Pontin}, David I. and {Priest}, Eric R.},
        title = "{Magnetic reconnection: MHD theory and modelling}",
      journal = {Living Reviews in Solar Physics},
         year = 2022,
        month = dec,
       volume = {19},
       number = {1},
          eid = {1},
        pages = {1},
          doi = {10.1007/s41116-022-00032-9},
       adsurl = {https://ui.adsabs.harvard.edu/abs/2022LRSP...19....1P}
}

@ARTICLE{2019ApJ...870..114S,
       author = {{Sim{\~o}es}, Paulo J.~A. and {Reid}, Hamish A.~S. and {Milligan}, Ryan O. and {Fletcher}, Lyndsay},
        title = "{The Spectral Content of SDO/AIA 1600 and 1700 {\r{A}} Filters from Flare and Plage Observations}",
      journal = {\apj},
         year = 2019,
        month = jan,
       volume = {870},
       number = {2},
          eid = {114},
        pages = {114},
          doi = {10.3847/1538-4357/aaf28d},
archivePrefix = {arXiv},
       eprint = {1808.01488},
 primaryClass = {astro-ph.SR},
       adsurl = {https://ui.adsabs.harvard.edu/abs/2019ApJ...870..114S}
}

@ARTICLE{2024ApJ...973L..10Y,
       author = {{Yadav}, Rahul and {Kazachenko}, Maria D. and {Afanasyev}, Andrey N. and {Cauzzi}, Gianna and {Reardon}, Kevin},
        title = "{Photospheric Pore Rotation Associated with a C-class Flare from Spectropolarimetric Observations with DKIST}",
      journal = {\apjl},
         year = 2024,
        month = sep,
       volume = {973},
       number = {1},
          eid = {L10},
        pages = {L10},
          doi = {10.3847/2041-8213/ad758d},
archivePrefix = {arXiv},
       eprint = {2408.16956},
 primaryClass = {astro-ph.SR},
       adsurl = {https://ui.adsabs.harvard.edu/abs/2024ApJ...973L..10Y}
}

@ARTICLE{2019A&A...632A.112Y,
       author = {{Yadav}, Rahul and {de la Cruz Rodr{\'\i}guez}, Jaime and
         {D{\'\i}az Baso}, Carlos Jos{\'e} and {Prasad}, Avijeet and
         {Libbrecht}, Tine and {Robustini}, Carolina and
         {Asensio Ramos}, Andr{\'e}s},
        title = "{Three-dimensional magnetic field structure of a flux-emerging region in the solar atmosphere}",
      journal = {\aap},
         year = 2019,
        month = dec,
       volume = {632},
          eid = {A112},
        pages = {A112},
          doi = {10.1051/0004-6361/201936790},
archivePrefix = {arXiv},
       eprint = {1910.13279},
 primaryClass = {astro-ph.SR},
       adsurl = {https://ui.adsabs.harvard.edu/abs/2019A&A...632A.112Y}
}

@article{Uitenbroek2001,
	Adsurl = {http://adsabs.harvard.edu/abs/2001ApJ...557..389U},
	Author = {{Uitenbroek}, H.},
	Doi = {10.1086/321659},
	Journal = {\apj},
	Month = aug,
	Pages = {389-398},
	Title = {{Multilevel Radiative Transfer with Partial Frequency Redistribution}},
	Volume = 557,
	Year = 2001}

@ARTICLE{2010ApJ...709.1362L,
       author = {{Leenaarts}, J. and {Rutten}, R.~J. and {Reardon}, K. and {Carlsson}, M. and {Hansteen}, V.},
        title = "{The Quiet Solar Atmosphere Observed and Simulated in Na I D$_{1}$}",
      journal = {\apj},
         year = 2010,
        month = feb,
       volume = {709},
       number = {2},
        pages = {1362-1373},
          doi = {10.1088/0004-637X/709/2/1362},
archivePrefix = {arXiv},
       eprint = {0912.2206},
 primaryClass = {astro-ph.SR},
       adsurl = {https://ui.adsabs.harvard.edu/abs/2010ApJ...709.1362L}
}

@ARTICLE{2023ApJ...958...54Y,
       author = {{Yadav}, Rahul and {Kazachenko}, Maria D. and {Afanasyev}, Andrey N. and {de la Cruz Rodr{\'\i}guez}, Jaime and {Leenaarts}, Jorrit},
        title = "{Solar Atmospheric Heating Due to Small-scale Events in an Emerging Flux Region}",
      journal = {\apj},
         year = 2023,
        month = nov,
       volume = {958},
       number = {1},
          eid = {54},
        pages = {54},
          doi = {10.3847/1538-4357/acfd2b},
archivePrefix = {arXiv},
       eprint = {2309.06452},
 primaryClass = {astro-ph.SR},
       adsurl = {https://ui.adsabs.harvard.edu/abs/2023ApJ...958...54Y}
}

@ARTICLE{2020SoPh..295..172R,
       author = {{Rimmele}, Thomas R. and {Warner}, Mark and {Keil}, Stephen L. and {Goode}, Philip R. and {Kn{\"o}lker}, Michael and {Kuhn}, Jeffrey R. and {Rosner}, Robert R. and {McMullin}, Joseph P. and {Casini}, Roberto and {Lin}, Haosheng and {W{\"o}ger}, Friedrich and {von der L{\"u}he}, Oskar and {Tritschler}, Alexandra and {Davey}, Alisdair and {de Wijn}, Alfred and {Elmore}, David F. and {Fehlmann}, Andr{\'e} and {Harrington}, David M. and {Jaeggli}, Sarah A. and {Rast}, Mark P. and {Schad}, Thomas A. and {Schmidt}, Wolfgang and {Mathioudakis}, Mihalis and {Mickey}, Donald L. and {Anan}, Tetsu and {Beck}, Christian and {Marshall}, Heather K. and {Jeffers}, Paul F. and {Oschmann}, Jacobus M. and {Beard}, Andrew and {Berst}, David C. and {Cowan}, Bruce A. and {Craig}, Simon C. and {Cross}, Eric and {Cummings}, Bryan K. and {Donnelly}, Colleen and {de Vanssay}, Jean-Benoit and {Eigenbrot}, Arthur D. and {Ferayorni}, Andrew and {Foster}, Christopher and {Galapon}, Chriselle Ann and {Gedrites}, Christopher and {Gonzales}, Kerry and {Goodrich}, Bret D. and {Gregory}, Brian S. and {Guzman}, Stephanie S. and {Guzzo}, Stephen and {Hegwer}, Steve and {Hubbard}, Robert P. and {Hubbard}, John R. and {Johansson}, Erik M. and {Johnson}, Luke C. and {Liang}, Chen and {Liang}, Mary and {McQuillen}, Isaac and {Mayer}, Christopher and {Newman}, Karl and {Onodera}, Brialyn and {Phelps}, LeEllen and {Puentes}, Myles M. and {Richards}, Christopher and {Rimmele}, Lukas M. and {Sekulic}, Predrag and {Shimko}, Stephan R. and {Simison}, Brett E. and {Smith}, Brett and {Starman}, Erik and {Sueoka}, Stacey R. and {Summers}, Richard T. and {Szabo}, Aimee and {Szabo}, Louis and {Wampler}, Stephen B. and {Williams}, Timothy R. and {White}, Charles},
        title = "{The Daniel K. Inouye Solar Telescope - Observatory Overview}",
      journal = {\solphys},
         year = 2020,
        month = dec,
       volume = {295},
       number = {12},
          eid = {172},
        pages = {172},
          doi = {10.1007/s11207-020-01736-7},
       adsurl = {https://ui.adsabs.harvard.edu/abs/2020SoPh..295..172R}
}

@ARTICLE{2013ApJ...764...33D,
       author = {{de la Cruz Rodr{\'\i}guez}, J. and {Piskunov}, N.},
        title = "{DELO-Bezier Formal Solutions of the Polarized Radiative Transfer Equation}",
      journal = {\apj},
         year = 2013,
        month = feb,
       volume = {764},
       number = {1},
          eid = {33},
        pages = {33},
          doi = {10.1088/0004-637X/764/1/33},
archivePrefix = {arXiv},
       eprint = {1212.2737},
 primaryClass = {astro-ph.SR},
       adsurl = {https://ui.adsabs.harvard.edu/abs/2013ApJ...764...33D}
}

@ARTICLE{2017A&A...597A..16P,
       author = {{Piskunov}, Nikolai and {Valenti}, Jeff A.},
        title = "{Spectroscopy Made Easy: Evolution}",
      journal = {\aap},
         year = 2017,
        month = jan,
       volume = {597},
          eid = {A16},
        pages = {A16},
          doi = {10.1051/0004-6361/201629124},
archivePrefix = {arXiv},
       eprint = {1606.06073},
 primaryClass = {astro-ph.IM},
       adsurl = {https://ui.adsabs.harvard.edu/abs/2017A&A...597A..16P}
}

@book{2003isp..book.....D,
	Adsurl = {https://ui.adsabs.harvard.edu/abs/2003isp..book.....D},
	Author = {{del Toro Iniesta}, J.~C.},
	Booktitle = {Introduction to Spectropolarimetry, by Jose Carlos del Toro Iniesta, pp.~244.~ISBN 0521818273.~Cambridge, UK: Cambridge University Press, April 2003.},
	Month = apr,
	Pages = {244},
	Title = {{Introduction to Spectropolarimetry}},
	Year = 2003}

@ARTICLE{2019A&A...626A..33H,
       author = {{Hansteen}, V. and {Ortiz}, A. and {Archontis}, V. and {Carlsson}, M. and {Pereira}, T.~M.~D. and {Bj{\o}rgen}, J.~P.},
        title = "{Ellerman bombs and UV bursts: transient events in chromospheric current sheets}",
      journal = {\aap},
         year = 2019,
        month = jun,
       volume = {626},
          eid = {A33},
        pages = {A33},
          doi = {10.1051/0004-6361/201935376},
archivePrefix = {arXiv},
       eprint = {1904.11524},
 primaryClass = {astro-ph.SR},
       adsurl = {https://ui.adsabs.harvard.edu/abs/2019A&A...626A..33H}
}

@ARTICLE{2015ApJ...812...11V,
       author = {{Vissers}, G.~J.~M. and {Rouppe van der Voort}, L.~H.~M. and {Rutten}, R.~J. and {Carlsson}, M. and {De Pontieu}, B.},
        title = "{Ellerman Bombs at High Resolution. III. Simultaneous Observations with IRIS and SST}",
      journal = {\apj},
         year = 2015,
        month = oct,
       volume = {812},
       number = {1},
          eid = {11},
        pages = {11},
          doi = {10.1088/0004-637X/812/1/11},
archivePrefix = {arXiv},
       eprint = {1507.00435},
 primaryClass = {astro-ph.SR},
       adsurl = {https://ui.adsabs.harvard.edu/abs/2015ApJ...812...11V}
}

@article{Peter2014Sci,
	Adsurl = {https://ui.adsabs.harvard.edu/abs/2014Sci...346C.315P},
	Archiveprefix = {arXiv},
	Author = {{Peter}, H. and {Tian}, H. and {Curdt}, W. and {Schmit}, D. and {Innes}, D. and {De Pontieu}, B. and {Lemen}, J. and {Title}, A. and {Boerner}, P. and {Hurlburt}, N. and {Tarbell}, T.~D. and {Wuelser}, J.~P. and {Mart{\'\i}nez-Sykora}, Juan and {Kleint}, L. and {Golub}, L. and {McKillop}, S. and {Reeves}, K.~K. and {Saar}, S. and {Testa}, P. and {Kankelborg}, C. and {Jaeggli}, S. and {Carlsson}, M. and {Hansteen}, V.},
	Doi = {10.1126/science.1255726},
	Eid = {1255726},
	Eprint = {1410.5842},
	Journal = {Science},
	Month = {Oct},
	Number = {6207},
	Pages = {1255726},
	Primaryclass = {astro-ph.SR},
	Title = {{Hot explosions in the cool atmosphere of the Sun}},
	Volume = {346},
	Year = {2014}}

@ARTICLE{2023MNRAS.521.3882K,
       author = {{Kaithakkal}, Anjali J. and {Borrero}, J.~M. and {Yabar}, A. Pastor and {de la Cruz Rodr{\'\i}guez}, J.},
        title = "{A reconnection-driven magnetic flux cancellation and a quiet Sun Ellerman bomb}",
      journal = {\mnras},
         year = 2023,
        month = may,
       volume = {521},
       number = {3},
        pages = {3882-3897},
          doi = {10.1093/mnras/stad759},
archivePrefix = {arXiv},
       eprint = {2303.06331},
 primaryClass = {astro-ph.SR},
       adsurl = {https://ui.adsabs.harvard.edu/abs/2023MNRAS.521.3882K}
}

@ARTICLE{Socas2006,
       author = {{Socas-Navarro}, Hector and {Mart{\'\i}nez Pillet}, Valent{\'\i}n and
         {Elmore}, David and {Pietarila}, Anna and {Lites}, Bruce W. and
         {Manso Sainz}, Rafael},
        title = "{Spectro-Polarimetric Observations and Non-Lte Modeling of Ellerman Bombs}",
      journal = {\solphys},
         year = 2006,
        month = may,
       volume = {235},
       number = {1-2},
        pages = {75-86},
          doi = {10.1007/s11207-006-0049-x},
archivePrefix = {arXiv},
       eprint = {astro-ph/0508667},
 primaryClass = {astro-ph},
       adsurl = {https://ui.adsabs.harvard.edu/abs/2006SoPh..235...75S}
}

@ARTICLE{2017A&A...598A..33L,
       author = {{Libbrecht}, Tine and {Joshi}, Jayant and
         {Rodr{\'\i}guez}, Jaime de la Cruz and {Leenaarts}, Jorrit and
         {Ramos}, Andr{\'e}s Asensio},
        title = "{Observations of Ellerman bomb emission features in He I D$_{3}$ and He I 10 830 {\r{A}}}",
      journal = {\aap},
         year = 2017,
        month = feb,
       volume = {598},
          eid = {A33},
        pages = {A33},
          doi = {10.1051/0004-6361/201629266},
archivePrefix = {arXiv},
       eprint = {1610.01321},
 primaryClass = {astro-ph.SR},
       adsurl = {https://ui.adsabs.harvard.edu/abs/2017A&A...598A..33L}
}

@ARTICLE{2004ApJ...614.1099P,
       author = {{Pariat}, E. and {Aulanier}, G. and {Schmieder}, B. and {Georgoulis}, M.~K. and {Rust}, D.~M. and {Bernasconi}, P.~N.},
        title = "{Resistive Emergence of Undulatory Flux Tubes}",
      journal = {\apj},
         year = 2004,
        month = oct,
       volume = {614},
       number = {2},
        pages = {1099-1112},
          doi = {10.1086/423891},
       adsurl = {https://ui.adsabs.harvard.edu/abs/2004ApJ...614.1099P}
}

\appendix
\section{Ellerman Bomb Identification Method}
\label{app:EBdetection}
\begin{figure}
    \centering
    \includegraphics[width=1\linewidth]{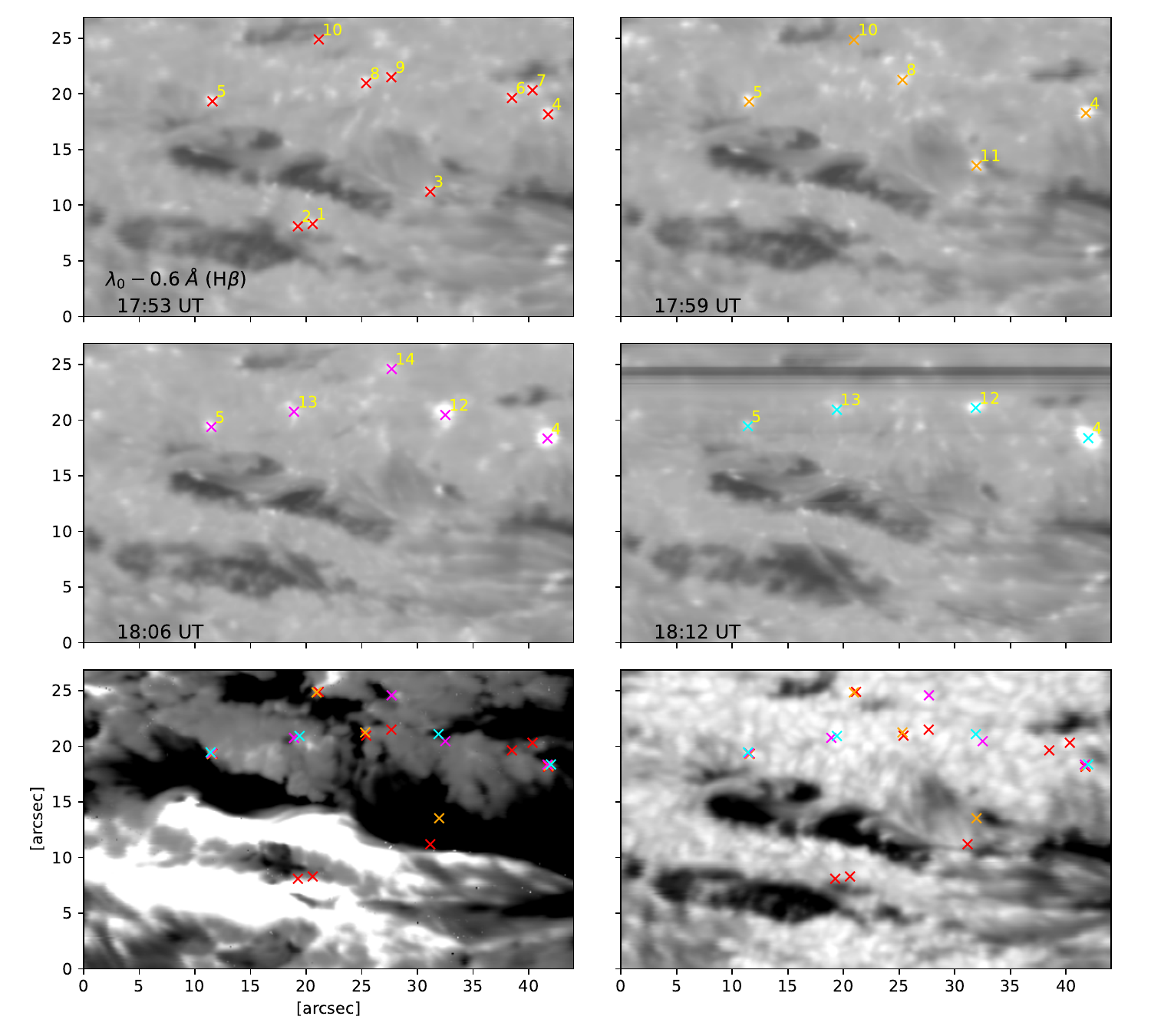}
    \caption{EB identification using intensity-threshold criteria. Top four panels: ViSP/DKIST H$\beta$ blue-wing intensity images showing the identified EBs, marked by crosses (`$\times$') and labeled accordingly. Different `$\times$' colors refer to different raster maps. Bottom left panel: Line-of-sight magnetic field inferred from the photospheric ViSP Fe~I~6302~\AA~line, where white and black represent positive and negative magnetic polarities, respectively. The magnetic field is saturated at $\pm 500$~G. The crosses indicate the locations of the identified EBs within the magnetic region. Bottom right panel: Same as the bottom left panel, but showing the continuum intensity. }
    \label{fig:idenitfy}
\end{figure}

Ellerman bombs (EBs) were identified based on their characteristic intensity enhancement in the wings of the H$\beta$ line. Our EB detection method consists of two stages of intensity threshold criteria. The initial identification was performed by selecting pixels that exceeded a specified intensity threshold in the averaged H$\beta$ blue/red wings at $\pm$ 0.65~\AA~ from the line center. 
After testing several intensity thresholds and spectral positions, we found that an intensity threshold of 0.75 at $\pm$0.65~\AA~H$\beta$ wings effectively identified EB-like features, although not all of the initially detected candidates were EBs. In the final stage, all candidates were manually inspected for the characteristic EB spectral signature, which is enhanced H$\beta$ wings without a significant enhancement of the line core. The candidates were classified as EB if their wing contrast in H$\beta$, defined as $I_{\mathrm{ctr}} =(I_{\mathrm{EB},\lambda} -I_{\mathrm{QS},\lambda})/I_{\mathrm{QS},\lambda}$, exceeded 20$\%$, where $I_{\mathrm{EB}}$ and $I_{\rm QS}$ represent the EB and the average quiet-Sun intensities, respectively. This additional criterion reduces contamination from other chromospheric brightenings and network bright points. Moreover, EBs having more than 4 pixels considered for further analysis. We note that very faint EBs exist below this threshold; however, most of these features do not exhibit a clear characteristic of EBs.  Using these criteria, we identified a total of 24 EB detections across four ViSP raster scans, corresponding to 14 unique EBs. Of these, two EBs were detected at similar locations in all four raster scans, while four EBs were detected in two consecutive raster scans.
The identified EBs are highlighted in Figure~\ref{fig:idenitfy}. 

\section{Sensitivity of the \ion{Fe}{1} and \ion{Na}{1}~D1 lines to Lower Solar Atmosphere}
\label{sec:responsefunction}
The sensitivity of a spectral line is commonly evaluated using response functions (RFs, \citealt{2003isp..book.....D}), which provide information on the atmospheric layers where the Stokes profiles respond to perturbations in the physical parameters of the model atmosphere. We used the inferred model atmosphere in the EB region from STiC inversions to compute the response functions of Stokes $I$ to temperature. 

\begin{figure}
    \centering
    \includegraphics[clip,trim=.4cm 0.5cm 0.4cm 0.5cm,width=0.7\linewidth]{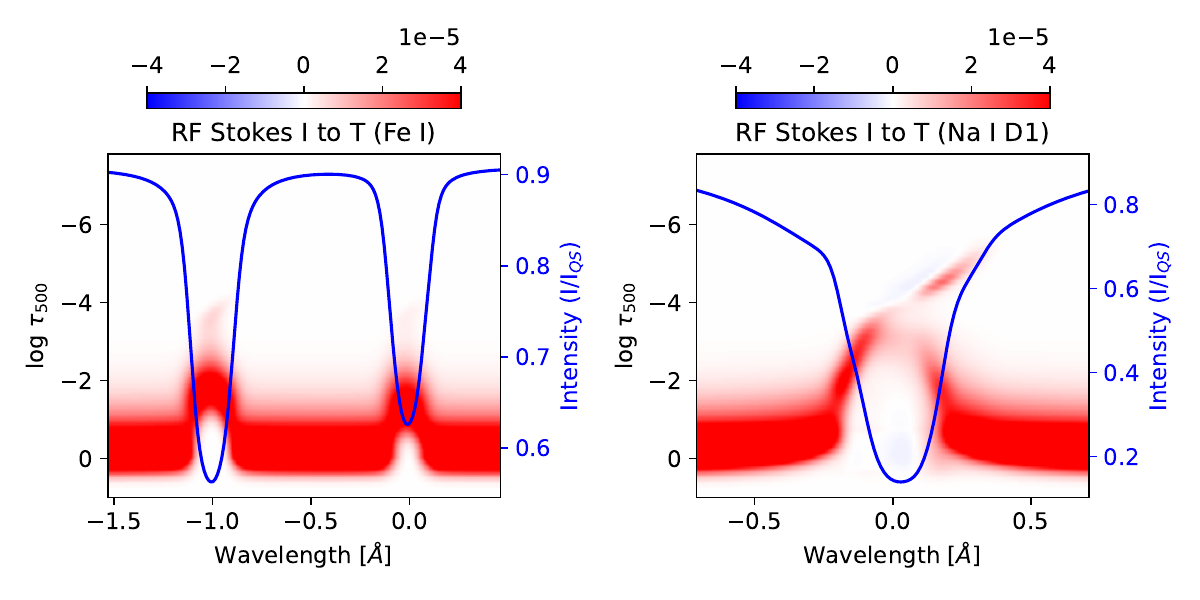}
    \caption{Response functions of the \ion{Fe}{1} 6301 line pair (left) and the \ion{Na}{1}~D1 line (right) to temperature as functions of optical depth and wavelength at the EB location. The blue profiles represent the observed intensity. The color bar in each panel shows the magnitude of response functions.}
    \label{fig:RF}
\end{figure}

Figure~\ref{fig:RF} presents the RF of the \ion{Na}{1}~D1 and \ion{Fe}{1} spectral lines to temperature perturbations at the EB location. 
From these plots it is evident that the \ion{Fe}{1} line pair mainly responds to temperature variations in the photosphere, around log~$\tau_{500}\sim0$ to $-2$, while the \ion{Na}{1}~D1 line shows sensitivity extending to higher atmospheric layers, reaching the upper photosphere and lower chromosphere (log~$\tau_{500}\sim-4$). The \ion{Fe}{1} line is strongly sensitive to the photosphere, whereas the core is weakly sensitive to the upper photosphere (log~$\tau_{500}\sim-3$), and at this location the wings and core of the \ion{Na}{1} line are stronger. Together, the two lines provide complementary diagnostics of the atmospheric layers where EBs are most likely to form, enabling a reliable reconstruction of the thermal structure associated with EB features. The response functions further show that the inferred atmospheric parameters are well constrained below log~$\tau_{500}\sim-4$, while the reliability of the inversion decreases at higher layers due to the reduced sensitivity of the spectral lines.

\end{document}